\documentclass[sigconf]{acmart} %
\AtBeginDocument{%
  }
\newcommand*\paperwebsite{https://5gdescrambler.github.io}
\newcommand*\permanentartifact{https://doi.org/10.5281/zenodo.22641598}

\usepackage{makecell}
\usepackage{subcaption}
\usepackage{multirow}
\usepackage[linesnumbered,ruled]{algorithm2e}

\usepackage{csquotes}
\usepackage{tabularx}
\usepackage{ragged2e}
\usepackage{tikz}
\newcommand*\emptycirc[1][1ex]{\tikz\draw (0,0) circle (#1);}
\newcommand*\halfcirc[1][1ex]{
  \begin{tikzpicture}
  \draw[fill] (0,0)-- (90:#1) arc (90:270:#1) -- cycle ;
  \draw (0,0) circle (#1);
  \end{tikzpicture}}
\newcommand*\fullcirc[1][1ex]{\tikz\fill (0,0) circle (#1);}

\makeatletter
\newenvironment{minipagealg}
{
	\par\noindent
	\begin{minipage}{\columnwidth}
		\begingroup
		\let\@latex@error\@gobbletwo
		\begin{algorithm}[H]
		}
		{
		\end{algorithm}
		\endgroup
	\end{minipage}
	\par
}
\makeatother

\newenvironment{DIFnomarkup}{}{}

\copyrightyear{2026}
\acmYear{2026}
\setcopyright{cc}
\setcctype{by}
\acmConference[CCS '26]{Proceedings of the 2026 ACM SIGSAC Conference on Computer and Communications Security}{November 15--19, 2026}{The Hague, Netherlands}
\acmBooktitle{Proceedings of the 2026 ACM SIGSAC Conference on Computer and Communications Security (CCS '26), November 15--19, 2026, The Hague, Netherlands}
\acmDOI{10.1145/3830454.3846809}
\acmISBN{979-8-4007-2871-6/2026/11}

\begin{document}

\typeout{Current Penalies:^^J
\string\clubpenalty=\the\clubpenalty^^J
\string\widowpenalty=\the\widowpenalty^^J
\string\displaywidowpenalty=\the\displaywidowpenalty}%

\title[5GDescrambler]{5GDescrambler: Locating, Descrambling, and Decoding 5G Scheduling Information (long version)}

\author{Fritz Windisch}
\affiliation{
	\department{KASTEL Security Research Labs}
	\institution{Karlsruhe Institute of Technology}
	\city{Kalrsuhe}
	\country{Germany}}
\email{fritz.windisch@kit.edu}

\author{Thorsten Strufe}
\affiliation{
	\department{KASTEL Security Research Labs}
	\institution{Karlsruhe Institute of Technology}
	\city{Kalrsuhe}
	\country{Germany}}
\email{thorsten.strufe@kit.edu}

\begin{abstract} %
  Tracking users in 5G NR has recently been successfully demonstrated by exploiting various side-channels.
  This allows for identification of individuals, classification of user activity in real time as well as tracking by fingerprinting, affecting billions of users with a 5G subscription and companies with private 5G deployments.
  However, previous work relies on weak operator configurations that leak networking parameters--either the radio network temporary identifier (RNTI) or scrambling factor ($N_{ID}$) during handshake--or to inefficiently brute-force Downlink Control Information (DCI).

  In this paper we present a novel technique exploiting algebraic structure to reverse DCI scrambling, fully integrated into an open-source end-to-end binary DCI sniffing pipeline.
  It provides enabling input for subsequent attacks like live tracking of users and supports automatic detection of control channel configurations used.
  We demonstrate the robustness and performance of our approach with measurement campaigns against deployments of srsRAN, OpenAirInterface5G, and two commercial vendors.
  It reaches block error rates of less than $1\%$ at SNRs below expected values for efficient communication, while performing significantly faster on a reference sample than a previously suggested passive technique brute-forcing the required parameters. In addition, it is entirely passive and does not rely on any side-channel leakage.
\end{abstract}



\maketitle

\newpage
\section{Introduction}
5G is the fifth generation mobile communication standard, serving billions of users worldwide \cite{EricssonMobilityReport}. It offers ultra-reliable and low latency communications, enhanced mobile broadband and massive machine-type communications to operators and users, enabling new use cases in various sectors such as healthcare, industry, robotics, autonomous vehicles, and edge computing.

5G was designed with more security and privacy in mind than prior standards.
As such, the concept of concealing the subscriber permanent identifier using a subscription concealed identifier, and better authentication measures like the 5G Authentication and Key Agreement (5G-AKA) protocol were introduced. Subsequent studies have demonstrated, however, that impersonating other users \cite{basinFormalAnalysis5G2018}, linking user identities across sessions \cite{chlosta5GSUCIcatchersStill2021}, and replay attacks \cite{cremersComponentBasedFormalAnalysis2019} are still possible under specific circumstances.

Additionally, compared to LTE, the secrecy of scheduling information was improved in 5G by introducing scrambling using UE-specific parameters as a radio performance feature to control channels such as the physical downlink control channel (PDCCH), now requiring the knowledge of parameters communicated over encrypted Radio Resource Control (RRC) signaling for decoding \cite[Section 7.3.2.3]{ts38_211}. Previously in LTE, only the well-known cell id $N_{ID}^{cell}$ and time position were required to decode PDCCH, greatly simplifying the process for passive sniffers. Even though scrambling is not a cryptographically secure primitive in the traditional sense, it inadvertently complicates the recovery of information when parameters used during encoding are not known, and recovering parameters or encoded contents directly is considered to be difficult by related work \cite{luoSni5GectPracticalApproach, ludant5GSniffingHarvesting2023}. While intercepting initial configuration parameters is possible due to lack of encryption during initial cell attachment, further configurations via RRC are mandated to be exchanged in encrypted form as soon as 5G radio security (access stratum security) is enabled, where legally possible \cite[Section 5.3.2]{ts33_501}, leading to changes in parameters no longer being visible to observers.
Learning the scheduling information transmitted on PDCCH can be exploited to record accurate traces of users participating in that cell, similar to identified dumps of encrypted traffic.
Such traces can be used by malicious parties to conduct traffic analysis revealing the users' identity \cite{ludant5GSniffingHarvesting2023}, webpages visited \cite{kohlsLostTrafficEncryption2019, website_fingerprinting}, apps used \cite{wangWhatAppAppUsage}, and videos watched \cite{baeWatchingWatchersPractical, video_fingerprinting}, or, generally, linking pseudonymous sessions across time and space.

On PDCCH, Downlink Control Information (DCI) is used to notify the user equipment (UE) of downlink transmissions or uplink grants issued by the 5G base station (gNB), containing all required information for localization, transmission scheme and expected size. The scrambling introduced on PDCCH is generated using two parameters that the recipient UE learns during cell association, and can subsequently be updated using encrypted RRC: its Radio Network Temporary Identifier (RNTI) and scrambling factor (PDCCH DM-RS scrambling ID). Recent studies have established that inverting the scrambling on the PDCCH to learn UE-specific parameters is possible but requires brute-forcing up to $2^{44}$ combinations using polar decoding operations \cite{ludant5GSniffingHarvesting2023}, which is time-consuming and thus not practical for real-time attacks \cite{wanNRScopePractical5G2024}.
Therefore some later approaches rely on using prior-knowledge, side-channel information or optimized brute-force \cite{ludant5GSniffingHarvesting2023, wanNRScopePractical5G2024, luoSni5GectPracticalApproach, gardnerEfficientMethodologyDeAnonymize2020} for sniffing the PDCCH of a 5G cell and to provide traces.

Without any insider knowledge from the system or eavesdropping on the cell association of a UE, a passive eavesdropper has to overcome three fundamental challenges:\\
\textit{C1) Unknown configuration parameters: }UEs are notified by the gNB about search spaces and hence know where to expect DCI messages on the resource grid, a configuration not known to a passive eavesdropper. %
Apart from initial configuration, PDCCH candidate configuration parameters later established during reconfiguration via RRC are commonly encrypted and, again, unknown to external parties.\\
\textit{C2) Unpredictability of channel: }Knowing the scrambling factor is  necessary to read and correlate demodulation reference symbols (DMRS), for channel amplitude and phase estimation. Apart from initial parameters, they are conveyed to a UE over encrypted RRC-signaling, and must be discovered by a passive eavesdropper through other means.\\

\textit{C3) DCI descrambling: }DCI are scrambled by a sequence generated from two 16-bit parameters, only known to gNB and UE: RNTI and scrambling factor. Furthermore decoding differs based on the decoded length of the DCI payload, which can range from 12 to 140 bits. %
Apart from initial configuration, the UE receives these parameters through encrypted RRC-signaling. To sniff, an adversary needs to find a way to obtain these parameters as efficient as possible. Prior work relied on brute-forcing the entire value space with one polar decoding operation per attempt, which is expensive to perform, or on side-channels providing this knowledge by other means.\\

In this work, we present a fully passive 5G PDCCH sniffer that, to the best of our knowledge, is the first to address all of the aforementioned challenges before, but also after enabling access stratum (AS) security, without resorting to brute-force. Our approach includes a novel technique that leverages structural properties of the PDCCH scrambling and DCI encoding to bypass scrambling without requiring prior knowledge, side channels or brute force. Since the technique only depends on the DCI decoding chain, compared to previous works in the field, specification-level changes are required for mitigation of the underlying algebraic properties. Our 5G PDCCH sniffer outputs successfully decoded binary DCI with their assignment to the flow of an individual RNTI, transmission timestamps and direction (uplink/downlink). We therefore provide enabling input to attacks from other works, such as live traffic classification in 5G cells, breaching the privacy of users on the cell, and allowing for tracking of users \cite{ludant5GSniffingHarvesting2023, wangWhatAppAppUsage, kohlsLostTrafficEncryption2019, baeWatchingWatchersPractical}.

For sniffing PDCCH using our solution, we consider a fully passive adversary that only observes over-the-air transmissions and has no legitimate access to the target cell.

We report true positive rates above $99\%$ at $SNR \ge 10 dB$ and a block error rate of less than $1\%$ at $SNR \ge 6.5 dB$ while decoding DCI candidates, at speeds of up to 300$\times$ faster than prior work, as we demonstrate on a comparable sample with tuned thresholds, so the prior work decodes all contained DCI candidates successfully. In our evaluation testbeds the aforementioned radio conditions were shown to be within operating conditions achievable using off-the-shelf SDR hardware in radio range of a gNB. These results demonstrate that our sniffer implementation and novel descrambling technique enable robust and performant binary DCI decoding.
As part of our work, we implemented an entire 5G synchronization and binary DCI decoding pipeline utilizing our technique in Rust, which we publish open source as part of this work\footnote{All artifacts (source code, build instructions, synthetic 5G RAN captures and configuration examples) are available at \url{\paperwebsite} and \url{\permanentartifact}.}.\\

Our contributions are as follows:
\begin{itemize}
    \item We present a novel approach to bypass scrambling entirely on PDCCH by exploiting the structural properties of PDCCH scrambling and DCI encoding.
    \item We develop a technique to infer the configuration of PDCCH without prior knowledge or active participation by the attacker even after access stratum (AS) security is enabled.
    \item We provide a channel estimator that does not rely on any reference symbols.
    \item We implement our techniques as an open-source end-to-end binary DCI sniffer called \textit{5GDescrambler} and evaluate our techniques on real-world samples captured from srsRAN, OpenAirInterface5G, [vendor 1 redacted] and [vendor 2 redacted]. It offers extraction of binary DCI linked to the pseudonymous RNTI of respective UEs, providing traffic analysis tools and subsequent attacks with timestamps, traffic direction (uplink/downlink) and received binary DCI of the transmitted traffic through various interoperable output formats. It does not perform further post-processing tasks, such as traffic size estimation, encrypted uplink/downlink decoding, UE-state reconstruction, flow linking, traffic classification or user tracking, but aims to facilitate these attacks for future work.
    \item We propose and discuss defense strategies against prior and our approach, and responsibly disclose our findings.
\end{itemize}

\textbf{Structure:} We first provide 5G background, assumptions, our novel PDCCH decoding technique bypassing scrambling, and the design of \textit{5GDescrambler}. We then evaluate robustness and performance, discuss related work, and conclude with limitations and future work.

\section{Background}
In this section we are going to describe the necessary background required for the construction of a passive 5G sniffer, visiting some physical layer, cell synchronization and resource scheduling details.

\subsection{5G NR Physical layer}
In 5G radio access networks (RAN), user equipments (UE, e.g. a mobile phone) communicate with the base station (gNB). Like LTE, 5G NR relies on orthogonal frequency division multiplexing (OFDM) with cyclic prefix (CP) for transmission of a time-frequency grid of symbols in time domain and subcarriers in frequency domain \cite{ts38_211}.

5G groups symbols into hierarchical time-domain structures, where frames (10ms) are divided into subframes, slots and symbols. The exact structure depends on the numerology and frequency band in use and is known by UEs and passive observers alike.
In frequency, grouping is also performed. One subcarrier at a specific point in time is called a resource element (RE). 12 subcarriers in frequency each form resource blocks (RB), of which one RB at one symbol in time is referred to as a resource element group (REG).

The resulting resource grid can then be used for data transmission.
A passive observer can obtain the transmitted grid and all required parameters by performing cell synchronization, which is explained in the following.

\subsection{5G NR Cell synchronization}
\label{sec:sync}
Synchronization in 5G NR is performed by first synchronizing time using a primary and secondary synchronization sequence (PSS \& SSS), before obtaining initial configuration and cell parameters such as the cell id $N_{ID}^{cell}$ from the repeatedly sent physical broadcast channel (PBCH) and system information block 1 (SIB1) \cite{ts38_211, ts38_213}. 

The UE can then attach to the cell using the physical random access channel (PRACH). The PRACH message exchange ends with the gNB resolving contention (resolves identifier collisions across UEs), and providing the UE with its cell radio network temporary identifier (C-RNTI) and initial configuration. The entire handshake can be observed passively, deducing the conveyed information. The UE can then proceed with radio resource control (RRC), during which access stratum (AS) security should be setup, mandated by TS 33.501 \cite[Section 5.3.2]{ts33_501} where legally possible, encrypting all subsequent RRC-signaling. Using encrypted RRC messages, the UE can then be reconfigured to receive a new C-RNTI mid-operation and adapt configurations without requiring reattachment, possible in multiple ways e.g. by using \textit{rach-LessHO} introduced by Release 18 (see TS 38.331 \cite[Section 5.3.5.3, 5.3.5.5.1 and 5.3.5.5.2]{ts38_331}), concealing the C-RNTI and new configurations from observers of the initial RACH handshake, making further tracking challenging.

\subsection{5G NR Resource Scheduling}
\label{sec:background_resource_scheduling}

Since many UEs share a cell in 5G, resource scheduling is required to coordinate UEs that communicate simultaneously.
This is performed by using the physical downlink control channel (PDCCH), transmitting uplink and downlink grants to the UE, which are also referred to as Downlink Control Information (DCI). For scheduling, the DCI contains the location and size in frequency and time domain in the grid, as well as the modulation and coding scheme (MCS) to use. Transmission size depends on the allocation size and MCS in use, which in 5G uses quadrature amplitude modulation (QAM) from 4-QAM (QPSK) to 1024-QAM.

To locate DCI on PDCCH, the UE is informed by the gNB about control resource sets (CORESET) and search spaces it is supposed to monitor. CORESET configurations are highly customizable and specify the PDCCH layout in frequency domain. As such, it provides the positions of control channel elements (CCE) and specifies the interleaving of DCI resource element groups (REG) through a CCE-to-REG mapping. Furthermore the placement of CORESET can be altered by specifying regions through a frequency resource assignment bitmap. Additionally a CORESET specifies the starting symbol within a slot and the duration of the CORESET in symbols ($n_{sym} \in \{1, 2, 3\}$). Search space configurations specify the amount of candidates to attempt to decode from a given CORESET configuration per aggregation layer (either 1, 2, 4, 8, or 16 CCEs per candidate), alongside their presence in slots. While UEs receive these configurations from the gNB, passive observers can only obtain initial configurations from SIB1 and those intercepted during RACH. Therefore finding CORESET and search space configurations, given their customizability, is more challenging in this case.

When attempting to decode a candidate from a search space, the UE has to determine the configuration of the DCI and whether the DCI is intended for itself. Therefore, in 5G encoded DCI are XOR-ed with a scrambling gold sequence using the RNTI and a scrambling factor, and the checksum for the decoded DCI is XOR-ed with the RNTI, both leading to checksum errors if the hypothesis of parameters tested was wrong. The UE can thus attempt decoding all possible hypotheses and upon receiving a checksum error determine that this transmission was either flawed, had a different recipient, or used different parameters than tested. This strategy is also referred to as blind decoding. For a passive observer this entails a challenge though: Since, in generality, RNTI and scrambling factor are not known beforehand, the scrambling sequence can not be constructed and reverted using XOR, leading to the observer not being able to decode PDCCH transmissions.

\subsection{5G NR PDCCH Transport Process}

\begin{figure}
    \centering
    \includegraphics[width=0.85\linewidth]{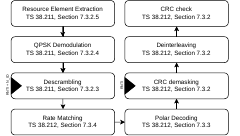}
    \caption{Steps to decode DCI in 5G NR. Descrambling and CRC demasking require access to the RNTI and the scrambling factor $\boldsymbol{N_{ID}}$.}
    \label{fig:dci_decode}
\end{figure}

In order to decode DCI, the UE has to perform the following steps (see Figure \ref{fig:dci_decode}):

\textbf{Resource Element Extraction} The UE needs to locate the REs belonging to the DCI candidate (TS 38.211 \cite[Section 7.3.2.5]{ts38_211}).

\textbf{QPSK Demodulation} The UE performs QPSK demodulation on the located REs (TS 38.211 \cite[Section 7.3.2.4]{ts38_211}), resulting in a sequence of logarithmic likelihood ratios (LLR). The amount of LLR values is denoted by $E$.

\textbf{Descrambling} The UE needs to descramble the DCI by XOR-ing a scrambling sequence to the input data (TS 38.211 \cite[Section 7.3.2.3]{ts38_211}). XOR-ing in the case of LLRs can be performed by flipping the sign of the respective value accordingly. The scrambling sequence is the gold sequence given in \cite[Section 5.2.1]{ts38_211} initialized by \cite[Section 7.3.2.3]{ts38_211}:

\begin{align}
\label{math:cinit}
    c_{init} = \left(n_{RNTI} * 2^{16} + n_{ID}\right) \mod 2^{31}
\end{align}

The two 16-bit inputs to this sequence depend on which kind of search space is used and whether a scrambling factor was configured using the RRC parameter \textit{pdcch-DMRS-scramblingID}.
In a common search space (CSS) or when \textit{pdcch-DMRS-scramblingID} is not set, $n_{RNTI}$ is set to $0$ and $n_{ID}$ is set to the physical-layer cell identity $N_{ID}^{cell}$, which is obtained from PSS/SSS during cell synchronization.
Otherwise, in a UE-specific search space (USS) with set \textit{pdcch-DMRS-scramblingID}, $n_{RNTI}$ is set to the C-RNTI obtained by the UE during synchronization and $n_{ID}$ is set to the \textit{pdcch-DMRS-scramblingID}. We will subsequently refer to the former case as ``shared'' and the latter as ``ue-specific''. Furthermore we will refer to $n_{ID}$ as the scrambling factor $N_{ID}$.

Notably, while providing de-correlation for blind-decoding between DCI for different UEs, this step also introduces a significant challenge for sniffing efforts in 5G.
While it is unclear to the authors whether this effect was intentionally introduced by 3GPP, recent work in the field mentions this as a major challenge for sniffers in 5G \cite{luoSni5GectPracticalApproach, ludant5GSniffingHarvesting2023}, since without access to the initialization parameters, which can be sent via encrypted channels, descrambling could not be performed efficiently.

\textbf{Rate Matching} The $E$ LLR values obtained after descrambling need to be rate matched to the polar coding matrix size, denoted by $N$ (TS 38.212 \cite[Section 7.3.4]{ts38_212}). Depending on the values for $E$ and $N$ different rate matching modes apply. If $E >= N$, repetition was used while encoding. The UE can thus sum repeated values in this case to improve confidence in the repeated LLRs. If $E < N$, not all values could be sent by the gNB, which used either puncturing (removing the bits with least reliable indexes) or shortening (removing the last bits of the transmission). Puncturing is used if $K / E \leq 7/16$, where $K$ is the amount of decoded DCI bits plus the CRC size. Else shortening is used.

\textbf{Polar decoding} The UE applies a polar decoder, for example a checksum-aided successive cancellation list (CA-SCL) polar decoder, to decode the input vector (TS 38.212 \cite[Section 7.3.3]{ts38_212}). Some bits in the output vector, called frozen bits, are deliberately inserted zero bits with known positions, enabling error correction.

\textbf{Deinterleaving} The $N$ resulting bits are deinterleaved to $K$ output bits (TS 38.212 \cite[Section 7.3.2]{ts38_212}). Interleaving is applied to DCI and checksum bits in order to distribute information across the sent bits. This step also removes the frozen bits inserted during interleaving, resulting in $A$ DCI bits and $24$ CRC-bits.

\textbf{CRC demasking} The last 16 CRC bits of the decoded CRC are XOR-ed with the RNTI used for the DCI, to distinguish the recipient (TS 38.212 \cite[Section 7.3.2]{ts38_212}).

\textbf{CRC check} The CRC of the DCI bits is computed and compared to the given CRC (TS 38.212 \cite[Section 7.3.2]{ts38_212}). If it matches, CRC is removed and the DCI bits are forwarded. Else, the candidate is discarded.\\

Given the challenge that RNTI and $N_{ID}$ are, apart from initial configuration before enabling encryption on RRC, not known to a passive observer, we will now continue by presenting the design of \textit{5GDescrambler}, including our novel approach to decode DCI despite being scrambled.

\section{Design of 5GDescrambler}
In this section we will describe our assumptions, followed by the introduction of our technique to bypass scrambling on 5G PDCCH, and a description of the components of \textit{5GDescrambler}.
\subsection{Assumptions}
Our assumptions are as follows:

\textbf{Network model} We assume the network to consist of one gNB and multiple UEs communicating over the air using 5G NR. All features of 5G NR may be used, including techniques like beamforming.

\textbf{Attacker model} We assume the adversary to be a fully passive observer, limited to executing efficient algorithms. She is in possession of a computational device (e.g. a sufficiently recent COTS laptop) and an SDR able to capture at the frequency of the target cell, while receiving at the bandwidth of the target cell.
She will never become active on the network, including association with the network, or injecting any other messages.
She will hence remain entirely undetectable by provider and subscribers.
When beamforming is used by the gNB, she has to be positioned in a similar angle to the gNB as the legitimate UE and within transmission range of the gNB, to ensure that she is physically able to overhear the corresponding PDCCH.
Otherwise, the attacker can be positioned anywhere within the covered area of the cell.

\subsection{Bypassing scrambling on 5G PDCCH}
\label{sec:technique}
Our technique exploiting algebraic structure to decode 5G DCI despite scrambling with unknown parameters being present is based on the observation, that all steps of the encoding chain are linear, and, as we prove by exhaustive analysis, that they are also reversible. We describe our approach in 4 parts: 1) We show that DCI encoding can be represented as a matrix multiplication of a vector consisting of all required parameters with a generator matrix $G$, 2) We show that $G$ is reversible to $G^{-1}$ enabling DCI descrambling, 3) We provide a strategy for error correction using linear equation systems and 4) give an intuition of complexity.

\textbf{1) Construction of generator matrix $\boldsymbol{G}$} We begin by showing that all steps in encoding DCI, given constant values for $K$ (DCI payload size + 24 CRC bits), $N$ (polar coding kernel size) and $E$ (encoded DCI size), can be attributed to one of seven possible base operations: matrix multiplication, XOR (denoted by $\oplus$), bit interleaving, vector concatenation (denoted by $\Vert$), zero-bit insertion, and bit repetition or deletion.

For the individual steps, this mapping can be performed as follows: For CRC attachment the generator polynomial $g_{CRC24C}$ can be converted to a CRC generation matrix $G_{CRC}$, which, given an input vector of 24-bits all set to $1$, produces the 24 CRC bits when applied using matrix multiplication. The 24 generated CRC bits are then appended to the input via vector concatenation. For CRC masking, the attached CRC can be XOR-ed with the RNTI. Subsequently bits are interleaved and frozen bits are inserted (zero bit insertion), leading to a vector size of $N$. During polar coding, the resulting vector is then multiplied by the predefined polar coding matrix $G_{N}$. Rate matching has three different modes and is handled either via bit repetition for the repetition mode, or via bit deletion for the puncturing and shortening modes. The result is the conversion of the vector of size $N$ to size $E$. Note, that the overall process is complex but well described in TS 38.212 \cite[Section 7.3.4 and 5.4.1]{ts38_212}.

Afterwards, scrambling needs to be computed and to be applied to the vector using XOR. The scrambling sequence $c(n)$ of size $E$ is defined as follows (TS 38.211 \cite[Section 5.2.1]{ts38_211}):
\begin{align*}
  c(n) &\equiv (x_1(n+1600) + x_2(n+1600)) \pmod 2
\end{align*}
where we initialize $x_1(0) = 1$ and $x_1(n)=0$ for $n \in [1;30]$ and $x_2(n) = c_{init}(n)$ for $n \in [0;30]$  with $c_{init}(n)$ denoting the $n$th bit of $c_{init}$ (see Equation \ref{math:cinit}) and the rest are obtained recursively with
\begin{align*}
  x_1(n+31) &\equiv (x_1(n+3) + x_1(n)) \pmod 2 \\
  x_2(n+31) &\equiv (x_2(n+3) + x_2(n+2) + x_2(n+1) + x_2(n)) \pmod 2
\end{align*}
for all $n$. Replacing the addition modulo two with XOR, we observe that the sequence consists of only XOR operations, making it fully linear. Furthermore we notice that the sequence $c(n)$ at any given position $n$ is always built from a distinct set of previous bits in both sequences $x_1(n)$ and $x_2(n)$ which are invoked by $c(n)$. Assuming a fixed position $n$ in the sequence $c(n)$, we can thus formulate positional scrambling functions $c_n(c_{init})$ providing the resulting scrambling sequence $c$ at concrete position $n$ for a given $c_{init}$. The positional scrambling functions $c_n(c_{init})$ can be constructed by repeatedly mapping input bits of $x_1$ and $x_2$ to the XOR sum of their predecessors, ending expansion when reaching input bit positions below 31. When reaching bit positions below 31, the values will be known per definition for $x_1$ or by bits in $c_{init}$ for $x_2$. We then eliminate all operands mapping to $0$ in $x_1$ and simplify the equation so every operand appears zero or one times, since operands cancel out under XOR. We thus obtain the positional scrambling function $c_n(c_{init})$, consisting of an XOR sum of selected bits in $c_{init}$ and $x_1(0)=1$.
Our input is subsequently represented as a vector:
\[v_{init} = (1, c_{init}(0),c_{init}(1),\dots{},c_{init}(30)).\]

The positional scrambling function $c_n(c_{init})$ for bit position $n$ can be rewritten as an incidence vector $v_n$ in $\operatorname{GF}(2)$, where each contained operand is represented as a $1$ at the index of the operand. When multiplying $v_{init}$ with $v_{n}$, we thus compute the result of the positional scrambling function $c_n(c_{init})$, resulting in the scrambling bit at bit position $n$ of scrambling sequence $c$. By stacking $v_n$ for all $n \in [0; E[$, we then create a sequence generation matrix $M$, used to build the entire scrambling sequence $c$ of all $E$ bit positions at once by multiplying $v_{init}$ with $M$:
\begin{align*}
    M &= \begin{bmatrix}
            v_0 & v_1 & \cdots& v_{E-1}
        \end{bmatrix}\\
    c &= v_{init} M.
\end{align*}

Scrambling can thus also be rewritten as a matrix multiplication. Please note that in common search spaces, the RNTI is set to zero during scrambling. Matrices conforming to this scrambling behavior can be constructed analogous.

The final two steps concerning QPSK modulation and resource element extraction are not relevant for our analysis, since they can be reversed at will.

As a second building block towards our approach, we show that all base operations of DCI encoding steps can be merged into one matrix multiplication with a generator matrix $G$. For each step, we have a fixed original input vector $a$ and a previously performed step as matrix $X$. For every step, $X$ is rebuilt to encompass all previous as well as the current step.

Every XOR operation with a vector can be represented by appending a $1$ to the input vector and a row to matrix $X$:
\begin{align*}
        \begin{pmatrix}
            a_0& \cdots& a_n & 1
        \end{pmatrix}
        \begin{pmatrix}
            X_{0}^{0}& \cdots& X_{n}^{0}\\
            \vdots & \ddots & \vdots \\
            X_{0}^{m}& \cdots& X_{n}^{m}\\
            b_0& \cdots& b_n
        \end{pmatrix}
        = (a  X) \oplus b.
    \end{align*}

In case two steps should be merged via XOR, we can rewrite this to concatenate the respective input vectors $a$ and $b$ horizontally and previous generator matrices $X$ and $Y$ vertically as follows:
\begin{align*}
        \begin{pmatrix}
            a & b
        \end{pmatrix}
        \begin{pmatrix}
            X\\
            Y
        \end{pmatrix}
        = (a X) \oplus (b Y).
    \end{align*}

Trivially, bit interleaving, zero-bit insertion, bit repetition and deletion can be performed by performing the same permutation on the columns of matrix $X$, since this has identical impact on the output.

Lastly, concatenation of the output of two steps represented by matrices $X$ and $Y$ depending on the same input vector can be concatenated by stacking them horizontally:
\begin{align*}
        a
        \begin{pmatrix}
            X & Y
        \end{pmatrix}
        = (a X) \Vert (a Y).
    \end{align*}
Following these findings, we construct our generator matrix $G$ (see Algorithm \ref{alg:gen_matrix} in the supplementary material for reference) and respectively define our input vector $u$ to contain the following information:
\begin{align*}
u =
        \begin{pmatrix}
            CRC, DCI, 1, RNTI, N_{ID}, Padding
        \end{pmatrix}
    \end{align*}
where $CRC$ bits are all $1$\footnote{The CRC generation matrix $G_{CRC24C}$ expects an array of ones to indicate that no CRC mismatch has occurred. CRC mismatches are visible as zeros when performing decoding later, allowing for syndrome solving and error detection.}, $DCI$ are the DCI bits, $1$ is a fixed bit necessary for negation and required by some operations\footnote{In a linear context, without having a fixed $1$ available, no fixed/static negation is possible, since we are limited to matrix multiplication and concatenation while constructing matrices. Furthermore mechanisms like the scrambling sequence initialization require a fixed $1$ in their initialization. We therefore simply provide one as part of the input vector as a solution to this problem.}, while $RNTI$ and $N_{ID}$ are in their respective bit representations. 
In order to have further parity check bits available we can also append a series of padding zero-bits to the input vector. We will describe their origin later when discussing reversibility.

\textbf{2) Reversibility of G} Showing reversibility of $G$ is performed by exhaustion. Since the construction of $G$ depends solely on the input parameters of $K$ (DCI payload size + 24 CRC bits), $N$ (polar coding kernel size), $E$ (encoded DCI size) and whether common search space (CSS) or UE-specific search space (USS) is used, we can assert reversibility of $G$ for all possible combinations. Parameter $N$ can be derived from $K$ and $E$, eliminating this factor from possible combinations. A full analysis of the rank and nullity of all existing configurations is included in Table \ref{tab:mat_rank} in the supplementary material. We observe that apart from higher $K$ on aggregation layer 1 ($E = 108$), all $G$ have rank $56 + A$ in UE-specific search space and $41 + A$ in common search space respectively. We generally wish to extract the DCI, RNTI and $N_{ID}$, resulting in $A + 32$ bits to be extracted. So while $G$ is not full rank, for all but some higher $K$ configurations on aggregation layer 1, enough information is retained to be able to correctly and uniquely revert a multiplication of a vector with $G$.

Given $E$ corresponding to aggregation layers $2$, $4$, $8$ or $16$, $K$ ranges from 36 to 164, which is imposed by the minimum DCI size and maximum supported interleaver size during CRC interleaving. For all combinations in UE-specific search space, when computing $G^{-1}$, only one bit is independent, which corresponds to the most significant RNTI bit $MSB_{RNTI}$. This is reasonable, since only the 15 least significant RNTI bits are included in scrambling, not constraining $MSB_{RNTI}$ in the coding scheme. In practice, a mismatching $MSB_{RNTI}$ leads to a fixed syndrome in CRC, which is trivially detected and corrected. In common search space, usually $16$ bits are independent, since no RNTI bits are constrained by scrambling. This however also leads to a syndrome in CRC, which can subsequently be corrected by solving a linear equation system.

Given $E$ corresponding to aggregation layer $1$, $K$ ranges from 36 to 107, given restrictions of minimum DCI size and $K<E$. However, an upper bound can trivially be deducted, as there are multiple possible input combinations to achieve identical output bits\footnote{Not all combinations of output bits are valid, therefore this is an upper bound.} when \begin{align}
    \label{math:constraint}
    K > E - |RNTI| - |N_{ID}| = 108 - 16 - 16 = 76
\end{align}
where $|RNTI|$ and $|N_{ID}|$ are the amount of bits in RNTI and $N_{ID}$. Therefore, $G$ only has identical reversibility properties as in higher aggregation layers, when $K \le 76$. Given $K > 76$, possible inputs can be enumerated, but lead to false positives following the degree of freedom of $G$ not constrained by the coding scheme. This is a limitation of the standard however, also providing false positives to UEs initialized with parameters matching one of the possible solutions. Therefore, these configurations are to be avoided by operators in practical deployments to eliminate ambiguity.

Given the decoding matrix $G^{-1}$, an attacker can now obtain the vector $x$ from modulation and perform
\begin{align}
\label{math:reverse}
    x G^{-1} = u
\end{align}
where $u$ is the decoded vector containing the information before encoding, fixing independent RNTI bits by either matching or solving a unique syndrome in CRC, depending on configuration.

The rank of $G^{-1}$ is only dependent on the minimum of either polar kernel size $N$ or the encoded DCI size $E$. Since for aggregation layer $4$ and $8$ repetition is used, bits beyond the polar kernel size are repeated and therefore always linear dependent on other bits, no longer contributing to the rank of the matrix. This is also reflected in the nullity, which only contains repeated bits. Apart from this, nullity is $0$ and the rank equals the matrix dimensions for all configurations. Therefore the decoded vector $u$ now contains additional bits according to the rank of $G^{-1}$, which are all set to $0$ as padding-bits and can be used for further parity checks. These bits can be compared to frozen bits in polar coding, aiding error correction in selecting a correct candidate, which we will describe in the following.

\textbf{3) Error correction} In real-world communication channels, signal quality can degrade, leading to the reception of the originally transmitted vector $x$ convoluted by an error vector $e$, commonly denoted by $\hat{x} = x \oplus e$.

Given that $G^{-1}$ is a decoding matrix with restrictions in the output vector, error correction can be performed by solving a linear equation system (LES) in $\operatorname{GF}(2)$ (see Algorithm \ref{alg:err_correction} in the supplementary material for reference). Following the assumption that the smallest absolute LLR values in $\hat{x}$ correspond to the least reliable and thus most likely incorrect bits, we select them as the set $I$ and recompute them. Therefore, we assign a linear variable to every bit in $I$. We then form one linear equation per known bit in $u$, which are the CRC bits being $1$, the XOR operation bit being $1$, and padding bits being $0$. Each linear equation is constructed by taking a column from the decoding matrix $G^{-1}$, indicating which bits in $x$ are summed to compute a single bit in $u$, given by incidence in the column. For each bit in $x$ contributing to the current bit in $u$ (where incidence is $1$), we append it to the linear equation being built. We append either the corresponding linear variable for bits present in $I$, or append the actual value received in $\hat{x}$ for all other bits. The result of the linear equation should be the expected value of the bit in $u$, since the full decoding column for $u$ has been modeled.

By applying Gaussian elimination, we then solve for a binding of all selected likely incorrect bits $I$ that satisfies the expected values of the bits in $u$, applying their computed value. The nullity of the equation system varies, but is usually small. Bits that are free can not be corrected in $I$, in which case we skip them, not providing error correction capabilities for them. If the recomputed binding for the linear variables was correct, we obtain the originally transmitted vector $x$. Applying Equation \ref{math:reverse}, we receive $u$ despite errors being present.

\textbf{4) Complexity} When stating complexity we need to distinguish between offline matrix construction, error correction and decoding. Offline matrix construction can be performed fully ahead of time, executing the matrix generation algorithm and reversion to $G^{-1}$ once per existing configuration (1416 in total). The dimensions and properties of matrices involved can be seen in Table \ref{tab:mat_rank} in the supplementary material.

The error correction per decoding attempt can be reduced to Gaussian elimination in $\operatorname{GF}(2)$ with $|I|$ variables and $m$ equations, where $|I|$ denotes the number of bits to correct errors for and $m$ is the amount of known, fixed bits in $u$, consisting of the 24 CRC-bits, the XOR negation bit and all padding bits. Since error correction should be run in real-time on potentially thousands of DCI per second, choosing $|I|$, the amount of errors to correct, properly is vital. We thus later chose $10\%$ of bits in UE-specific search space and up to $5\%$ of bits in common search space as reasonable values in our experiments, selecting less bits in common search space, since the rank of the generator matrix is lower for common search spaces. These values can be configured freely for a trade-off between decoding robustness and performance however.

Lastly, decoding requires multiplication of the input vector $x$ with the decoding matrix $G^{-1}$. Afterwards, a syndrome needs to be corrected to fix and match the CRC. In UE specific search space, when only one bit is independent, this is performed by copying the expected syndrome directly from the generator matrix $G$. Else, if multiple bits are independent, the syndrome is corrected by solving a linear equation system. Complexity thus always includes multiplication of a vector of size $E$ with a matrix of the size of $E \times E$, before either copying a syndrome of size $K$ from $G$ or solving a linear equation system, which solves all the free bits from $G$ (see Table \ref{tab:mat_rank} in the supplementary material) to produce a correct syndrome for the CRC bits and the negation bit (33 total). The nullity of the LES is static per configuration and 0 for all configurations (apart from unusable configurations with $K \ge 99$ on AL 1). The solution is thus strictly defined.

Since we no longer require to brute-force the entire parameter space for $RNTI$ or $N_{ID}$, we do not have to attempt all possible $O(|RNTI|*|N_{ID}|)$ decoder executions, but can instead decode with one single execution in $O(1)$ (given in decoder executions).

\subsection{Components}
Our approach consists of four parts:

\textbf{1) Synchronization and preparation} In this step, the sniffer performs the normal 5G NR synchronization procedure (see section \ref{sec:sync}) continuously, obtaining time synchronization, public cell parameters and initial configuration as conveyed by PBCH and SIB1. It performs a multitude of digital signal processing steps to improve the quality of the resulting grid. Subsequently it performs channel estimation against all antennas used for sniffing and merges the resulting grids into one grid per frame, before sending the resulting frame grid including channel estimation information off to the next processing steps.

\textbf{2) Active group detection} Next we want to detect active REGs (subsequently referred to as groups) using QPSK in order to reduce the amount of processing required later and to obtain the estimated phase rotation of the respective group (see Algorithm \ref{alg:active_groups} in the supplementary material for reference).
To check each group on the grid for activity, we initially compare the 12 REs of the group to the information obtained by channel estimation from the previous step.
If not at least 9 amplitudes of REs are above a cutoff chosen from channel estimations, the group is considered inactive.
This already removes many groups from the grid that are not relevant to the operation of the sniffer (e.g. channel state information or PDSCH/PUSCH DMRS that all contain more than the accepted amount of REs set to zero).

Secondly we classify whether a group can plausibly use QPSK, with the goal of minimizing false negatives, which would potentially lead to missed DCI or configurations later.
This is done by estimating the angle deviation between constellation points as well as the difference in amplitude.
Since QPSK has one constellation point per quadrant, the standard deviation in angle and amplitude should be lowest for QPSK compared to higher QAM modulation schemes.
Of course we can not know before correlating against a reference signal which quadrant a point belongs to.
Therefore we sample the 12 REs of the group as complex numbers and mirror all to quadrant one. One issue remaining then is that angles clip at 0 and $\frac{\pi}{2}$.
Therefore we multiply by $4$ to spread the points to a full circle, before averaging angles and using a conjugated dot product to compare the angles to the average, resulting in a standard deviation of angles in the group.
For the standard deviation of amplitude, we simply take the amplitudes of the original complex numbers as a basis.

After computing both standard deviations for a group, we check them against a carefully chosen threshold of $< \frac{\pi}{3}$ for angle and $< 0.3 * avg_{amplitude}$ for amplitude, where $avg_{amplitude}$ is the average amplitude of the original complex numbers. Groups not matching the thresholds are considered not using QPSK, the rest of the groups are considered active.

All of the active groups in a symbol are then rotated equally so they match one of the four possible QPSK constellations. In order to not jump in phase between symbols, the rotation is also applied to all remaining symbols in a slot. A slot border is a natural boundary for PDCCH, since CORESETs can not span multiple slots. In case there is accumulating phase over subcarriers, it is also corrected.

Finally all active groups are grouped per frame and forwarded to the next processing step.

\textbf{3) DCI sniffing} Once our resource grid is prepared and active groups are determined we start to decode DCI (see Algorithm \ref{alg:sniffing_config} in the supplementary material for reference). If no configuration is known yet or was provided by the attacker (optional), we initialize with the common CORESET and search space configuration given by SIB1. Next, we iterate over all slots and potential candidates and check, whether all of their groups are denoted as active, before attempting to decode.

We originally only know the scrambled LLR sequence, aggregation layer and optionally the size of the decoded DCI, if provided with the candidate. Using this information we then apply our decoding approach as described in Section \ref{sec:technique} and return the decoded DCI, RNTI and $N_{ID}$, if successful. For every successfully decoded DCI we also mark the corresponding groups inactive, to not further process them in configuration finding. Sniffing can be parallelized per DCI candidate to decode, where performance is mainly dependent on the amount of active candidates to attempt to decode per slot and their aggregation layer, since higher aggregation layers take longer to decode.

Every DCI can consist of a limited amount of formats (e.g. 0\_0, 0\_1), discernible by their varying size. Therefore, their content can be parsed by exhaustive search over candidate DCI field interpretations, with decoding attempts on PDSCH/PUSCH for confirmation, determining the transmission location on the grid, the amount of information transmitted, as well as the transmission direction (uplink/downlink). We currently only parse the transmission direction, since we consider post-processing of binary DCI to be out of scope for our work, requiring significant development effort. The output of our sniffer can thus be used as input to subsequent fingerprinting attacks, such as identification and tracking of individuals \cite{ludant5GSniffingHarvesting2023}, and classification of user activity \cite{wangWhatAppAppUsage, kohlsLostTrafficEncryption2019, baeWatchingWatchersPractical}. %

\textbf{4) Configuration finding} With previously decoded DCI candidates being already removed, we now start a constrained incremental search for new DCI candidates (see Algorithm \ref{alg:sniffing_config} in the supplementary material for reference). Using brute-force to attempt to decode all possible combinations is hardly possible, since candidates can use almost arbitrary resource block (RB) positions on the grid due to extensive configuration options via interleaving patterns, CORESET resources and candidate index. This leads to a worst-case of ${N_{RB}}\choose{k}$ combinations, where $N_{RB}$ is the bandwidth used by the carrier in RBs and $k$ is the amount of RBs to choose.

However, since cell load usually varies over time, it is likely for newly detected active groups to contain one or more DCI candidates not known before. We therefore apply a strategy exploiting this characteristic. For every slot and possible symbol size of CORESETS ($n_{sym} \in \{1, 2, 3\}$), we collect a vector of active groups remaining after candidate decoding. For $n_{sym} \in \{2, 3\}$ we know, that only active groups can be used which are active in all symbols of the CORESET. Subsequently we start searching for new DCI candidates, which are checked by attempting to decode a selection of active groups in frequency domain with $n_{sym}$ groups in time domain. When decoding succeeds, we append the found candidate to the candidate list for further frames and return the decoded DCI.

The combinations our search tries, in order, are as follows: 1) For a fast-path, we check whether the amount of active groups detected exactly match one aggregation layer size. If so, we check the resulting candidate. 2) To catch all candidates that are simply offset in active groups without being interrupted by unrelated active groups, we try all sequences of active groups. 3) Exploiting that REG bundles, if not interrupted due to CORESET resource boundaries, always form minimum chunks of size $s$ of a fixed, known length per $n_{sym}$, we attempt to decode all combinations of chunks of size $s$. 4) Lastly we fall back to brute-force, to catch all configurations not covered by previous assumptions.

In order to balance performance and accuracy, we define a fixed amount of combinations to check per slot, $n_{sym}$ and aggregation layer. In case we do not catch a correct combination during one occurrence, it is likely that we will obtain it cheaper later when the cell is under a different amount of load. As more aggressive optimization towards performance, we also provide a \textit{fast mode}, consisting of only steps 1) and 2), or a \textit{super fast mode} consisting of only step 1). Using error correction while searching for candidate configurations is optional, to increase configuration finding speed under good channel conditions ($>= 20$dB SNR). Furthermore, configuration finding can be disabled after a specified amount of frames to transition to sniffing-only operation or disabled entirely, then requiring the attacker to provide configurations obtained by previous measurements.

Sniffing and configuration finding are concepts that need to be viewed separately. While disabling configuration finding may lead to the pipeline missing DCI, configurations also appear to be commonly reused between different UEs \cite{wanNRScopePractical5G2024} and to converge quickly in our experiments. Therefore, already found configurations can be cached and reused while disabling configuration finding after a set amount of frames or when no new configurations are found for some time.
It can however be required to re-run configuration finding after new configurations are deployed that have not been seen before. This can occur when new UEs enter the network, on handover, and RRC reconfiguration, including BWP switching, CORESET or search space updates for UEs. Then, unless configuration finding is re-enabled, we would miss affected DCI. Mobility does not affect configurations in use and thus has no effect on configuration completeness, unless candidates can not be read due to mobility induced artifacts. This only applies when the sniffer or the gNB change positions rapidly though, which in most cases should not apply.
We therefore aim to support pure sniffing in a real-time fashion, but tolerate configuration finding to not run in real-time, since configurations can be cached. Therefore, our full pipeline containing sniffing and configuration finding is only guaranteed to run in real-time, while configuration finding is disabled. Our experiments show however, that unless challenged with traffic of large and constant scheduler load, configuration finding also often performs in real-time.

\section{Implementation and Evaluation}
We subsequently implemented our design in Rust, providing extensive tooling to encode and decode DCI, perform cell search, decode waveforms from file or live via SDR, integration via packet capture files, tun/tan devices, a custom Wireshark dissector as well as JSON or human friendly output. We performed evaluation on the three challenges of C1) Unknown configuration parameters, C2) unpredictability of channel and C3) DCI descrambling, represented by their components of configuration finding, active group detection and DCI sniffing respectively.

\textit{Note, that there unfortunately is no state of the art we can directly compare against.}
The only prior approach with similar assumptions---lack of knowledge of configuration parameters from side-channels---is the \textit{5GSniffer} \cite{ludant5GSniffingHarvesting2023}.
We acquired the corresponding code to perform comparison measurements. With the exception of running it on the traces provided from the same source, however, it consistently crashed with segfaults.
The only sensible comparison possible hence was execution time of descrambling the given traces, and we can not provide robustness and performance results in the remaining experiments and plots.
We report on qualitative differences to prior art in our Related Work (Section \ref{sec:sota}), below.

\textbf{DCI descrambling:} For DCI descrambling (C3), we encoded DCI of various lengths to all aggregation layers, as well as matching random QPSK constellations.
We then applied our decoder with error correction enabled and without known parameters of $RNTI$ and $N_{ID}$, as well as a CA-SCL polar decoder using various list sizes $L$ and known parameters to the identical LLRs. We subsequently counted numbers of true positives (TP), true negatives (TN), false positives (FP) and false negatives (FN), computing block error rates (BLER, limited to valid DCI) and accuracy (for all classified data) for various SNRs.
We provide comparisons given different aggregation layers (maximum sane DCI input size for all), input lengths within aggregation layer 1 (known and unknown) and shared/UE-specific search spaces for all cases. Furthermore, we also compare our approach to polar decoding, expecting our approach to be less robust, since more information needs to be recovered, but similar in performance. For comparison with polar decoding, we mainly provided comparison with a list size of 8, since this appeared to yield the best computational performance versus decoding robustness trade-off in our case.
We generally expected common search spaces to yield lower robustness, since RNTIs are not used in scrambling initialization ($c_{init}$) and are thus not constrained, leading to only 8 bits remaining in the CRC check.
For aggregation layer comparison, we expected higher aggregation layers to yield higher robustness, due to the increased amount of redundancy.
Similarly we expected lower robustness for lower input sizes.
In the case of unknown input size, we expected our scheme to fail at $K>76$ due to the constraint established by formula \ref{math:constraint}.
Each experiment was conducted $10^4$ times for each SNR value.

\textbf{Active group detection:} For active group detection (C2) we wanted to determine the quality of detecting active resource blocks (RBs) using QPSK as modulation scheme, since those denote the relevant RBs for later analysis.
We therefore conducted an experiment providing our active group detector with arrays of inactive and active groups in a range of different SNRs.
We used various modulation schemes commonly applied in 5G for the active groups, ranging from QPSK to 256-QAM.
Each array was also rotated by a random value to determine channel estimation accuracy.
For each array we perform channel noise estimation based on a separate inactive group (identical to known inactive groups during cell synchronization) and classify the other group (either inactive or active), counting combinations of true positives (TP), true negatives (TN), false positives (FP) and false negatives (FN).
We then computed the F1-Macro score of the active group classifier to factor in accurate detection across both classes (active and inactive). We expected good recognition for QPSK and resistance against higher order modulation schemes.
For every group detected active we also determined the distance between estimated channel rotation and actual induced channel rotation, respecting the $\frac{\pi}{2}$ ambiguity of QPSK, implied by QPSK only providing one constellation point per quadrant.
We subsequently computed the standard deviation between subsequent runs, expecting our standard deviation to be significantly lower than random guessing (expected standard deviation of $\frac{\pi}{8}$).
For every SNR value we conducted the experiment over $10^6$ runs.

\textbf{Configuration finding and sniffing:} For evaluation of configuration finding (C1) and the full-chain pipeline (C1-C3), we tested \textit{5GDescrambler} against input samples captured from srsRAN, OpenAirInterface5G (OAI), [vendor 1 redacted] and [vendor 2 redacted].
The samples for srsRAN were taken from \textit{5GSniffer} \cite{ludant5GSniffingHarvesting2023}, providing a base-line for comparison. Their first sample is a recording of two UEs on a voice call, while their second sample contains one UE receiving occasional transmissions.
We captured the rest of the samples by placing a gNB, the UE(s) and the sniffer (using a USRP B210) in an equilateral triangle with edges of approximately 2.5 meters.
In case of OpenAirInterface5G, we publish two samples, one of a UE flooding the network and one of a UE sending continuous ping transmissions with our artifacts.
We collaborated with a national authority of IT security, which provided us with samples for [vendor 2 redacted] and [vendor 1 redacted], which we also publish with our artifacts.
For their setup, they also placed the UEs, sniffer and gNBs in an equilateral triangle of approximately 2.5 meters, using the \textit{Amarisoft LTE UE and NR simulator} to simulate 5 UEs flooding the cell, a USRP B210 SDR for the sniffer, and \textit{[vendor 1 redacted]} as well as \textit{[vendor 2 redacted]} family devices for the gNBs using 20MHz as cell bandwidth. All devices were placed in shielding boxes. For the [vendor 2 redacted] flood samples 2 and 3, shielding was removed and gain for the sniffer increased, resulting in large effects of noise in the samples, reducing SNR for these samples drastically.
For some samples, UEs were in connected state before the begin of the sample. This includes the OpenAirInterface5G samples and [vendor 2 redacted] flood 3.
Subsequently, we tested our \textit{5GDescrambler} on the samples, using configuration finding in \textit{super fast} and \textit{fast mode} with and without error correction, comparing the output with a reference provided as UE logs.
In case of the \textit{5GSniffer} samples, since no baseline was available, we compared with \textit{5GSniffer} output.
We measured the detection and decoding of candidates as the common tuple of TP, FP, and FN, to compute the resulting accuracy.
Furthermore, we measured the time taken for candidate decoding and finding on a reference PC\footnote{\label{fn:reference_machine}The reference PC is using a Ryzen 9750X3D and 5200MHz dual channel DDR5 RAM. Results were additionally verified on a Laptop using a Ryzen 7 PRO 7730U CPU.} and compare it to the base-line of the length of samples analyzed.
The capability to sniff traffic (not find configurations) live requires decoding times lower than sample length, which we expected to achieve.

\subsection{Results}
\begin{table*}[t]
\centering
\begin{tabular}{|l|r|r|r|r|r|r|r|r|r|r|r|r|c|}
\hline
\textbf{Waveform} & $SNR_{dB}$ & $TP$ & $FN$ & $FP$ & $I$ & $N_{ref}$ & $N_{dec}$ & $t_{ag}$ & $t_{candidate}^{dec}$ & $t_{candidate}^{find}$ & $t_{conv}^{cf}$ & $t_{budget}$ & $C_{corr}$\\
\hline
srsRAN \textit{5GSniffer} 1 & 32.73 & 431 & 0 & 0 & 49 & 480 & 480& 135.34ms & 65.73ms & 103.16ms & 1.47s & 4.97s & $\checkmark$ \\
srsRAN \textit{5GSniffer} 2 & 35.13 & 40 & 0 & 0 & 3 & 43 & 43 & 197.83ms & 71.20ms & 7.78ms & 3.07s & 4.99s & $\checkmark$ \\
OAI flood & 30.66 & 597 & 0 & 0 & 434 & 1031 & 1031 & 248.55ms & 102.43ms & 5.64ms & 0.04s & 4.99s & $\checkmark$ \\
OAI ping & 30.50 & 72 & 0 & 0 & 34 & 106 & 106 & 222.41ms & 72.84ms & 83.01ms & 0.83s & 4.99s & $\checkmark$ \\
{[vendor 1 redacted]} flood & 32.35 & 1789 & 0 & 0 & 13 & 1802 & 1802 & 320.48ms & 377.24ms & 9.28s & 3.67s & 4.99s & $\checkmark$ \\
{[vendor 2 redacted]} flood 1 & 29.59 & 4661 & 2 & 0 & 28 & 4691 & 4689 & 807.85ms & 2.20s & 2.16s & 2.83s & 4.66s & $\checkmark$ \\
{[vendor 2 redacted]} flood 2 & 13.64 & 3994 & 17 & 0 & 32 & 4043 & 4026 & 353.29ms & 974.38ms & 3.40s & 3.88s & 4.98s & $\checkmark$ \\
{[vendor 2 redacted]} flood 3 & 12.23 & 5291 & 28 & 0 & 40 & 5359 & 5331 & 367.68ms & 1.55s & 3.61s & 1.85s & 4.99s & $\checkmark$ \\
\hline
\end{tabular}

\caption{Results of combined configuration finding (super fast mode, disabled error correction) and sniffing (with error correction) on different samples.}{\begin{justify}$SNR_{dB}$: Average SNR measured at SSB; $TP$, $FN$, $FP$, $I$ (incorrectly decoded results); $N_{ref}$, $N_{dec}$: expected and decoded amount of DCI; $t_{ag}$, $t_{candidate}^{dec}$, $t_{candidate}^{find}$, $t_{conv}^{cf}$: time taken for active group detection, candidate decoding, candidate finding and for candidate finding to converge; $t_{budget}$: analyzed duration; $C_{corr}$: all candidates successfully found.\end{justify}}
\label{tab:res}
\end{table*}

\begin{figure}[t]
    \centering
    \begin{subfigure}{0.35\textwidth}
        \centering
        \includegraphics[width=\linewidth]{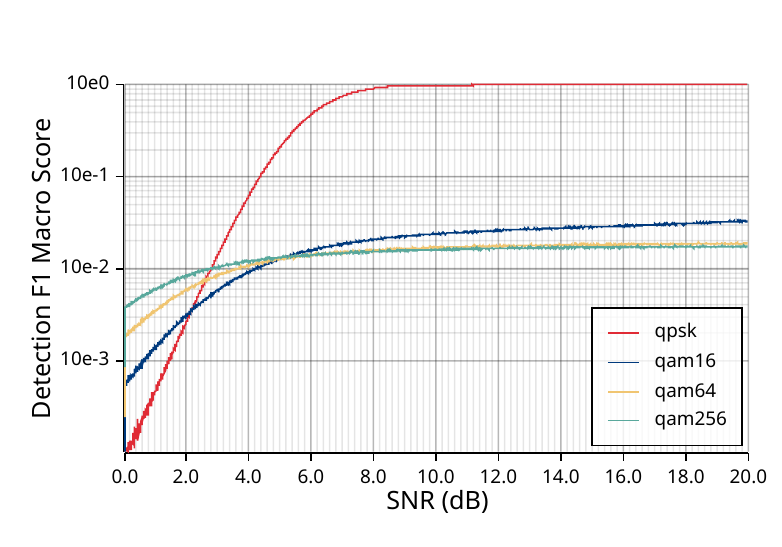}
        \caption{Active group detection F1-Macro score (log plot)}
        \label{fig:res_ag_det}
    \end{subfigure}
    \hfill
    \begin{subfigure}{0.35\textwidth}
        \centering
        \includegraphics[width=\linewidth]{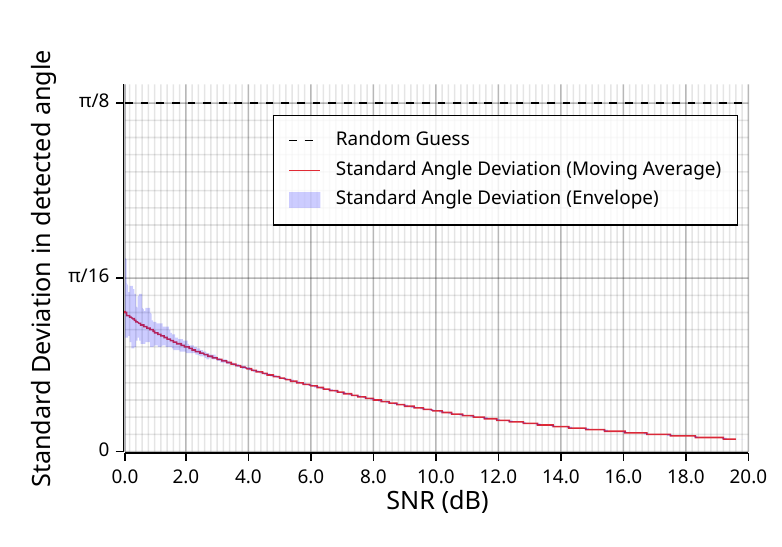}
        \caption{Active group std in detected angle}
        \label{fig:res_ag_angle}
    \end{subfigure}
    \caption{Results for active group detection
    }
    \label{fig:res_ag}
\end{figure}

\begin{figure}[t]
    \centering
    \begin{subfigure}{0.35\textwidth}
        \centering
        \includegraphics[width=\linewidth]{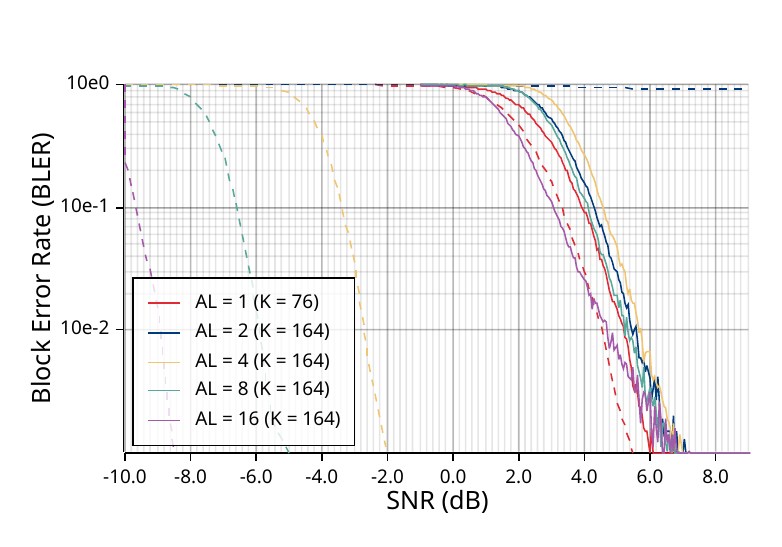}
        \caption{UE-specific Search Space (USS)}
        \label{fig:res_dci_al_ue}
    \end{subfigure}
    \hfill
    \begin{subfigure}{0.35\textwidth}
        \centering
        \includegraphics[width=\linewidth]{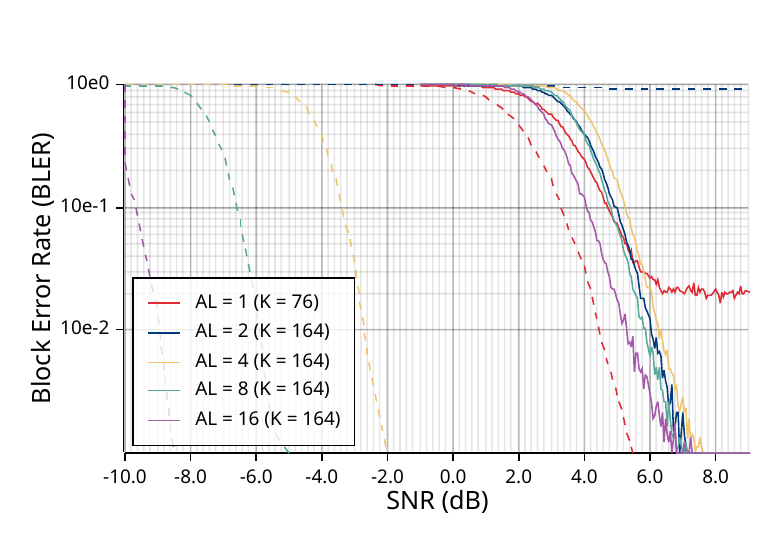}
        \caption{Common Search Space (CSS)}
        \label{fig:res_dci_al_cs}
    \end{subfigure}
    \caption{BLER for DCI descrambling and decoding per aggregation layer with maximum valid K. Our technique displayed as full line, CA-SCL polar decoding (L=8) as striped line.}
    \label{fig:res_dci_al}
\end{figure}

\begin{figure}[t]
    \centering
    \begin{subfigure}{0.35\textwidth}
        \centering
        \includegraphics[width=\linewidth]{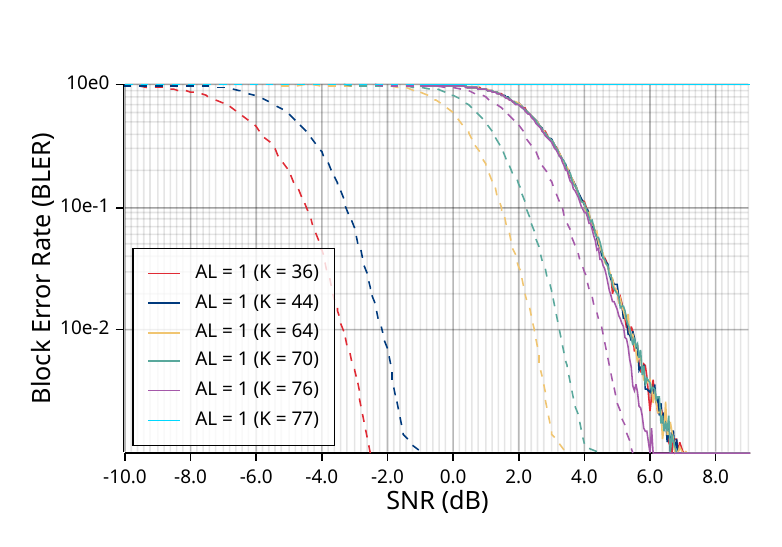}
        \caption{Automatic detection of K in UE-specific Search Space (USS)}
        \label{fig:res_dci_in_auto}
    \end{subfigure}
    \hfill
    \begin{subfigure}{0.35\textwidth}
        \centering
        \includegraphics[width=\linewidth]{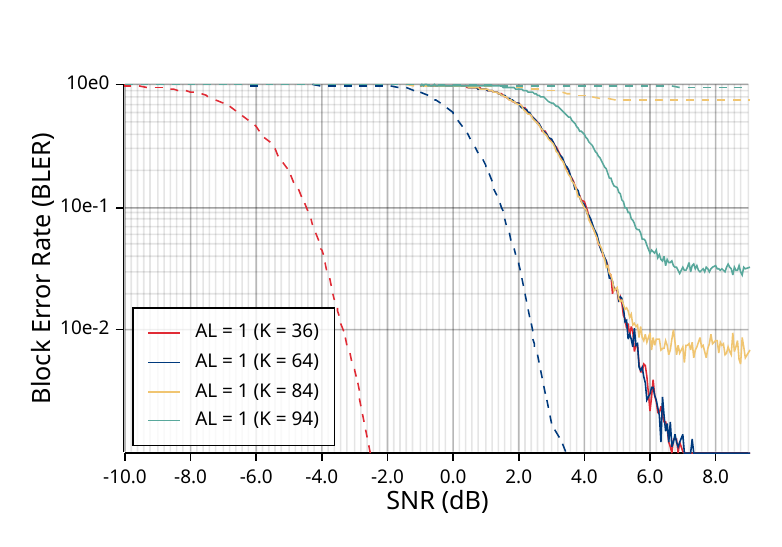}
        \caption{Known K in UE-specific Search Space (USS)}
        \label{fig:res_dci_in_fixed}
    \end{subfigure}
    \caption{BLER for DCI descrambling and decoding for different input sizes at aggregation layer 1. Our technique displayed as full line, CA-SCL polar decoding (L=8) as striped line.}
    \label{fig:res_dci_in}
\end{figure}

Overall, our results show that \textit{5GDescrambler} performs well under reasonable channel conditions.

For DCI descrambling (C3), it generally achieves excellent block error rates at reasonable SNRs.
When comparing aggregation layers with maximum valid DCI sizes, it performed best for aggregation layer 16 and worst for aggregation layer 4, with $\approx$1.5dB SNR difference to reach identical BLERs for both USS and CSS. For all other aggregation layers it performed equally well, and results are situated between the two extremes (see Figure \ref{fig:res_dci_al}).
This is surprising, since we initially expected better robustness for aggregation layers according to their rate of redundancy, which rises per layer.
In practice this is not relevant though, since above 8dB SNR all BLERs are below 0.1\% and thus negligibly low, showing that our choice to include the $|I|$ most likely incorrect bits (10\% in USS and 5\% in CSS) in the invalid bit correction set $I$ provides good results.
For CSS with aggregation layer 1 at maximum DCI size though, robustness stagnated at $\approx$1.3\% BLER, reaching an upper limit in accuracy due to missing constraints.
It's worthy of note that DCI in CSS at aggregation layer 1 are usually sent with input sizes of $\approx$63 to 69 bits (DCI format 0\_0 or 0\_1) where BLER also reaches values lower than 0.1\% beyond 8dB SNR.
This is considered poor channel quality in 5G, and can thus easily be obtained by the attacker, given the attacker can strategically position itself and use fine-tuned antenna setups.
As expected, for CSS robustness was $\approx$0.5dB worse than for USS.
Accuracy behaved as expected, since beyond 8dB SNR only spurious decoding failures occurred across all layers and configurations.

When comparing input sizes at aggregation layer 1 in USS (see Figure \ref{fig:res_dci_in}), we generally observed that BLERs were invariant to the input size until $K$ exceeded the theoretical bound of $76$ (see Equation \ref{math:constraint}).
When $K$ is not known, as previously mentioned, the algorithm attempts to decode with every possible size of $K$ from $K_{min}$ to $K_{max}$.
When reaching $K = 76$ in decoding, every possible input belongs to a valid decoded DCI, leading to the algorithm to terminate.
Therefore, no decoding attempts for higher $K$ can succeed, leading to a BLER of 1.
When constraining $K$ to a known value, beyond $K = 76$ block error rates and false positives increased steadily.
For CSS, the results are analogous, though robustness for CSS was $\approx$0.5dB worse and the same limitation concerning maximum DCI size on aggregation layer 1 applies as above.
This implies that for all realistic unambiguous configurations, \textit{5GDescrambler} will perform equally well, even at relatively poor channel conditions of 8dB SNR.

For a comparison of our technique to commonly used decoding techniques, we compare against a CA-SCL polar decoding implementation validated against Rob Maunder's 3GPP polar implementation \cite{robMaunder}.
When comparing robustness with the frequently used list size $L=8$ as a base-line, we measured significantly better results for most configurations at lower SNRs using polar decoding, especially in higher aggregation layers of 4 and above, since the amount of frozen bits in these layers is high. To further investigate error correction capabilities, we also performed controlled error injection on both implementations for individual bits at AL 1 and $K = 36$. The polar decoder tolerated single-bit errors on all tested input bits, while our technique only supports correcting errors within the most likely set of incorrect bits $I$ that are not free. Errors outside this scope lead to a decoding failure without subsequent recovery attempts, independent of $K$ and aggregation layer.

\begin{table}[t]
	\centering

	\begin{tabular}{|l|r|r|r|r|r|}
		\hline
		\textbf{AL} & \multicolumn{1}{c|}{CA-SCL} & \multicolumn{4}{c|}{\textit{5GDescrambler}} \\
		& \multicolumn{1}{c|}{polar} & \multicolumn{2}{c|}{err corr} & \multicolumn{2}{c|}{$\neg$err corr} \\
		& \multicolumn{1}{c|}{(L = 8)} & \multicolumn{1}{c|}{USS} & \multicolumn{1}{c|}{CSS} & \multicolumn{1}{c|}{USS} & \multicolumn{1}{c|}{CSS} \\
		\hline
		1 & 36.62 µs & 1.49 µs & 7.92 µs & 0.20 µs & 5.12 µs \\
		2 & 57.87 µs & 8.05 µs & 18.18 µs & 0.58 µs & 9.87 µs \\
		4 & 102.25 µs & 38.97 µs & 53.45 µs & 1.69 µs & 26.09 µs \\
		8 & 103.67 µs & 104.79 µs & 103.17 µs & 2.25 µs & 36.47 µs \\
		16 & 107.23 µs & 475.62 µs & 299.55 µs & 2.27 µs & 45.60 µs \\
		\hline
	\end{tabular}

	\caption{Comparison in decoding performance of single candidates between our CA-SCL polar decoder implementation (independent of search space type) and decoding by 5GDescrambler in USS and CSS using $K=36$.}
\label{tab:performance_vs_polar}
\end{table}

When comparing performance against our CA-SCL polar implementation with $L=8$, we do observe \textit{5GDescrambler} with error correction to outperform it significantly for aggregation layer 4 and lower, but match or fall behind on higher aggregation layers (see Table \ref{tab:performance_vs_polar}). Without error correction, \textit{5GDescrambler} outperforms the polar decoder significantly. However, depending on use case, various and also further optimized versions of polar decoding exist. Highly optimized polar decoding implementations such as Aff3ct \cite{aff3ct} can decode similar structures such as DCI in $\approx$10µs (AL = 1), $\approx$14µs (AL = 2|4), or $\approx$20µs (AL = 8|16) on our reference machine, outperforming \textit{5GDescrambler} for some scenarios, even though results are skewed due to Aff3ct not implementing the full DCI decoding chain.

We therefore do not claim our technique to be a replacement to conventional polar decoding, but instead view our approach as an efficient, sufficiently robust and reliable way to obtain the two unknown parameters and binary DCI data at the same time, even when channel conditions are as low as 8dB SNR. However, the error correction capabilities and silent decoding errors against controlled error injection remain unvalidated, since not all possible error patterns and configurations were investigated.

For active group detection (C2), we found that our classifier confidently detects active QPSK groups with an F1-macro score above 99\% for 10dB SNR or higher, which can reasonably be reached by attackers (see Figure \ref{fig:res_ag_det}).
At the same time, the classifier strongly rejected higher QAM modulation schemes as expected, approaching an F1-macro score of only about 3\%. %
Incorrectly classifying a few higher QAM groups as active only provides minor consequences, since it only slightly raises the complexity of finding configurations later.

Detection of rotated angles, invariant to QPSK rotations, became increasingly precise with higher SNRs (see Figure \ref{fig:res_ag_angle}).
At 10dB SNR we reached 8 times more precision in standard deviation than random guessing, following a logarithmic drop.
This precision is sufficient to establish correct constellations for the analyzed groups, since larger rotations occur naturally in the spectrum between individual symbols or between subcarriers, e.g., when transmitter and receiver clocks mismatch or due to the Doppler effect.

For combined testing of the full sniffer pipeline, we observed that \textit{5GDescrambler} produced minimal false positives or negatives on all samples (see Table \ref{tab:res}).
We observed that, especially for samples with high amounts of CSS DCI, it output invalid decoding results, though.
Upon closer inspection, all invalid decoded DCI were correct apart from the last bit of RNTI.
Interestingly, swapping all bits in a transmission in CSS, the last bit of the RNTI flips.
We could hence verify that the incurred errors were due to the sniffer interpreting those DCI with a 180 degree rotation.
This issue can be circumvented by either utilizing DMRS for channel estimation (as a side-channel) or CSI-RS to continuously estimate channel rotations with more accuracy.
While capturing traffic from multiple vendor gNB, such as OpenAirInterface5G, [vendor 1 redacted] and [vendor 2 redacted], we observed that RNTIs were often assigned randomly within a range of possible RNTIs, leading to the chance of a collision in the first 15 RNTI bits being low. Apart from this, all other DCI are decoded correctly. Since false positives, creating a new flow, or false negatives, corresponding to sub-percent packet rate drops, do not drastically change flow shapes, no negative effect is to be expected for traffic classification. For RNTI ambiguity, in case RNTIs collide often due to e.g. round-robin schemes being used in CSS \cite{rnti_management}, we recommend disambiguation by DMRS correlation, as later stated in our limitations.

\stepcounter{footnote}
\edef\snifferfootnoteone{\thefootnote}
	\begin{table*}[t]
		\centering
		\footnotesize
		\begin{tabularx}{\linewidth}{|l|*6{>{\hspace{0pt}\centering\arraybackslash}X}@{}|}
			\hline
			\textbf{Category} & \textbf{Fully passive} & \textbf{Post-RAR sniffing} & \textbf{Sniffing without brute-force} & \textbf{Configuration Finding} & \textbf{Supports all configurations} & \textbf{Specification-level changes required} \\
			\hline
			\textbf{LTESniffer}  \cite{hoangLTESnifferOpensourceLTE2023} & \checkmark & \checkmark & \checkmark & \checkmark & \checkmark & \checkmark \\
			\hline
			\makecell[lt]{\textbf{Syndrome descrambling}\\Gardner et al. \cite{gardnerEfficientMethodologyDeAnonymize2020}\\Flores et al. \cite{floresImplementationEvaluationSmart2023}\\Richards et al. \cite{richardsRNTIRecoveryOptimization2024}} & \checkmark & \checkmark & $\times$ & $\times$ & $\times$ Assume $N_{ID} = N_{ID}^{cell}$ and USS & $\times$ Use CSS or $N_{ID} \neq N_{ID}^{cell}$ \\
			\hline
			\makecell[lt]{\textbf{RACH side-channel}\\Wan et al. \cite{wanNRScopePractical5G2024}\\Luo et al. \cite{luoSni5GectPracticalApproach}\\Wang et al. \cite{wangSmartJammingDownlink2025}} & \checkmark & $\times$ & \checkmark & $-$ Indicated by SIB1 and RrcSetup & \checkmark & $\times$ Reconfigure UE post enabling AS security \\
			\hline
			\makecell[lt]{\textbf{Divide and conquer}\\Ludant et al. \cite{ludant5GSniffingHarvesting2023}} & $-$ Requires attacker to provide static configurations\footnotemark[\snifferfootnoteone] & $-$ Requires PDCCH configuration to be known\footnotemark[\snifferfootnoteone] & $\times$ & $\times$\footnotemark[\snifferfootnoteone] & \checkmark & $\times$ Randomize/ permute PDCCH config. per UE \\
			\hline
			\textbf{5GDescrambler} & \checkmark & \checkmark & \checkmark & \checkmark & \checkmark & \checkmark \\
			\hline
		\end{tabularx}
		\normalsize
		\caption{Comparison of our \textit{5GDescrambler} to related work.}{
			\begin{justify}
				\checkmark: Applies; $-$: Partially applies; $\times$: Does not apply; Fully passive: No attach/transmissions needed; Post-RAR sniffing: Works after updating configurations through RrcReconfiguration; Sniffing without brute-force: No brute-force of RNTI/$N_{ID}$; Configuration finding: Supports automatic PDCCH configuration detection. Supports all configuration: All possible PDCCH configurations supported; Specification-level changes required: Specification-level changes are required for the approach to be mitigated
			\end{justify}
		}
		\label{tab:related_work}
	\end{table*}
	\begin{DIFnomarkup}
	\footnotetext[\snifferfootnoteone]{While they provide a brute-force strategy on DMRS implemented in MATLAB, \textit{5GSniffer} \cite{ludant5GSniffingHarvesting2023} depends on configurations to be known beforehand.}
	\end{DIFnomarkup}

As shown in Table \ref{tab:res}, \textit{5GDescrambler} performed sniffing consistently in time that was lower than sample length, while configuration finding demonstrated real-time capability for all but one sample.
For srsRAN and OpenAirInterface5G samples, active group detection took below 5\% of available time (sample length), candidate decoding and candidate finding (in \textit{super fast mode} without error correction) each took below 3\% of time, while missing no candidates among the samples. For the \textit{[vendor 1 redacted] flood} and \textit{[vendor 2 redacted] flood 1} samples at high SNR, the sniffer successfully decoded all DCI for the former and missed 2 out of 4691 DCI ($\approx0.04\%$). Due to the [vendor 1 redacted] cell more commonly using QPSK for PDSCH, PDSCH is included within the active groups and needs to be analyzed by configuration finding, causing candidate finding to lose real-time capabilities on this sample. Large amounts of QPSK groups and high non time-varying scheduler load, as exhibited by flooding samples, constitutes a worst-case for configuration finding though. However, since candidate decoding in our current implementation waits for results from configuration finding for the previous slot, this also delays candidate decoding. During operation from SDR, this can cause queues in the pipeline to overflow on extended delays, resulting in skipped samples and corresponding DCI. For all other samples, including flooding samples from other vendors, configuration finding completes within sample time and is thus real-time capable. For samples where real-time capability by configuration finding can not be reached, it can be disabled after a set amount of frames. For the samples under test, the threshold when all configurations have been found as given by $t_{conv}^{cf}$ ranged from 0.04s (\textit{OAI flood}) to 3.88s (\textit{[vendor 2 redacted] flood 2}). Generally, for a specific vendor, the more DCI are scheduled within a sample, the faster configuration finding converges.
Candidate decoding behaved as expected, given it performs in real-time across all samples and is mainly dependent on aggregation layers in use and the amount of DCI candidates to attempt to decode. This can be seen when comparing the timings for the [vendor 1 redacted] sample (9 candidates with AL=4 in config), with the [vendor 2 redacted] samples (16 candidates with AL=1, 8 with AL=2, 4 with AL=4 and 3 with AL=8 in config). The [vendor 1 redacted] sample decodes faster than the [vendor 2 redacted] samples, since the amount of candidates is higher and AL 8 is used frequently by the [vendor 2 redacted] gNB. The two samples with lower SNR report some false negatives (at most $\approx0.5\%$), which at relatively low SNRs and in a flooding scenario is to be expected.
\textit{5GDescrambler} remains capable of live-sniffing across all samples with at most 48\% of available time (2.2s/4.66s) taken when using error correction, excluding delays induced by configuration finding.
Under good signal conditions, as given by most samples apart from \textit{[vendor 2 redacted] flood 2} and \textit{3}, this can be improved significantly by disabling error correction, leading to a reduction in processing time for the previous worst performing sample \textit{[vendor 2 redacted] flood 1} to only 242ms of time taken (below 5\% of the available time of 4660ms). The resulting output correctly represents time occurrence, identifiers, uplink/downlink classification as well as DCI contents, allowing for further post-processing to parse DCI contents, therefore providing enabling input to traffic fingerprinting attacks by future work.

We thus conclude that \textit{5GDescrambler} fulfilled expectations of sniffing 5G control channel traffic live with configuration finding disabled, while also performing configuration finding live for all but one sample, which is subject to the queue overflow limitation when processing from SDR (see Section \ref{sec:limitations}).

\section{Related Work}
\label{sec:sota}

\stepcounter{footnote}
\edef\snifferfootnotetwo{\thefootnote}
\stepcounter{footnote}
\edef\nrscopefootnote{\thefootnote}
\begin{table*}[t]
    \centering
    \footnotesize
    \begin{tabularx}{\linewidth}{|l|*{4}{>{\hspace{0pt}\centering\arraybackslash}X}@{}|*{3}{>{\hspace{0pt}\centering\arraybackslash}X}@{}|*{3}{>{\hspace{0pt}\centering\arraybackslash}X}@{}|*{2}{>{\hspace{0pt}\centering\arraybackslash}X}@{}|}
    	\hline
    	& \multicolumn{4}{c|}{\textbf{Binary PDCCH sniffing}} & \multicolumn{3}{c|}{\textbf{Post processing}} & \multicolumn{3}{c|}{\textbf{Attacks on privacy}} & \multicolumn{2}{c|}{\textbf{Evaluation details}} \\
        \hline
        \textbf{Paper} & \textbf{Configuration discovery} & \textbf{Efficient $RNTI$ and $N_{ID}$ recovery} & \textbf{Post-RAR Support} & \textbf{Binary DCI recovery} & \textbf{Parsed DCI recovery} & \textbf{Sched. message decoding} & \textbf{UE-state reconstruction} & \textbf{Flow linking} & \textbf{Traffic classification} & \textbf{Tracking} & \textbf{Continuous live operation} & \textbf{Oper. commer. network eval.} \\
        \hline
        LTESniffer \cite{hoangLTESnifferOpensourceLTE2023} & \fullcirc & \fullcirc & \fullcirc & \fullcirc & \fullcirc & \fullcirc & \fullcirc & \fullcirc & $\times$ & \fullcirc & \halfcirc & \fullcirc \\
        \hline
        5GSniffer \cite{ludant5GSniffingHarvesting2023} & \emptycirc\makebox[0pt][l]{\footnotemark[\snifferfootnoteone]} & \halfcirc\makebox[0pt][l]{\footnotemark[\snifferfootnotetwo]} & \fullcirc & \fullcirc & \halfcirc & $\times$ & $\times$ & $\times$ & \emptycirc & \fullcirc & \fullcirc & \fullcirc \\
        NR-Scope \cite{wanNRScopePractical5G2024} & \fullcirc & \fullcirc & $\times$ & \fullcirc & \fullcirc & $\times$\makebox[0pt][l]{\footnotemark[\nrscopefootnote]} & \fullcirc & \halfcirc & $\times$ & $\times$ & \fullcirc & \fullcirc \\
        Sni5Gect \cite{luoSni5GectPracticalApproach} & \fullcirc & \fullcirc & $\times$ & \fullcirc & \fullcirc & \fullcirc & \fullcirc & \fullcirc & $\times$ & \fullcirc & \fullcirc & $\times$ \\
        \hline
        5GDescrambler & \fullcirc & \fullcirc & \fullcirc & \fullcirc & \emptycirc & \emptycirc & $\times$ & $\times$ & $\times$ & $\times$ & \fullcirc & \emptycirc \\
        \hline
    \end{tabularx}
    \normalsize
    \caption{Capabilities of \textit{5GDescrambler} and related work.}{
    \fullcirc: Implemented and evaluated; \halfcirc: Partially implemented or evaluated; \emptycirc: Discussed but not implemented; $\times$: Unsupported}
    \label{tab:capabilities}
\end{table*}
\begin{DIFnomarkup}
\footnotetext[\snifferfootnotetwo]{\textit{5GSniffer} \cite{ludant5GSniffingHarvesting2023} implements an optimized brute-force strategy to recover the unknown UE-specific parameters of $RNTI$ and $N_{ID}$. Therefore, their approach is less efficient than approaches recovering the parameters from side-channels or by exploiting algebraic structure.}
\footnotetext[\nrscopefootnote]{\textit{NR-Scope} \cite{wanNRScopePractical5G2024} parses RRC messages, but only provides size estimations for the rest of scheduled messages.}
\end{DIFnomarkup}

Table \ref{tab:related_work} compares 5GDescrambler against existing passive 5G PDCCH sniffing approaches and the previous mobile network generation sniffer \textit{LTESniffer} \cite{hoangLTESnifferOpensourceLTE2023}. Although further active approaches like MITM attacks exist for 5G \cite{wangWhatAppAppUsage}, they were excluded as their set of challenges diverge drastically from passive sniffers (i.e. they do not need to circumvent scrambling).

In LTE, the set of challenges a PDCCH sniffer has to overcome are different from 5G. Since scrambling exists but only depends on the well-known cell id $N_{ID}^{cell}$ and slot number in a frame $n_s$, PDCCH can be decoded using conventional polar decoding without requiring the knowledge of additional UE-specific parameters. Therefore, previous generation sniffers such as \textit{LTESniffer} \cite{hoangLTESnifferOpensourceLTE2023} do not need to solve challenge 3 and thus do not require brute-forcing parameters or eavesdropping of RACH. Similarly, in LTE the cell-specific reference signal (CRS) accompanying PDCCH is generated using only well-known parameters, allowing for channel prediction without requiring a sniffer to solve challenge 2 (unpredictability of channel). However, RRC can also be encrypted in LTE, still requiring a sniffer to obtain configurations (challenge 1) from either RACH or by correlation.

Starting from 5G, all three challenges of unknown configuration parameters, unpredictability of channel and DCI descrambling need to be solved by PDCCH sniffers. In general, three different kinds of approaches for passive sniffing of PDCCH in 5G, partially solving the three challenges, currently exist:

\textbf{Syndrome descrambling} Syndrome descrambling approaches \cite{gardnerEfficientMethodologyDeAnonymize2020, floresImplementationEvaluationSmart2023, richardsRNTIRecoveryOptimization2024} consider the scrambling sequence as constant noise in polar decoding. Therefore, when not descrambling a DCI before decoding, any errors indicated by frozen bits being non-zero will form a fixed syndrome per unique scrambling sequence applied. A syndrome table is subsequently built, which provides a reverse lookup of syndrome bits to every possible scrambling sequence input (C-RNTI and $N_{ID}$). When a DCI candidate is received, the polar decoder can be run without descrambling, leading to a syndrome again. This syndrome can then be matched against the table.

These approaches suffer three issues though. First, syndromes are unique for every C-RNTI (16-bit), $N_{ID}$ (16-bit) and polar decoding combination (5 aggregation layers with $\sim$128 possible DCI lengths each), resulting in an unfeasible large search and storage space required\footnote{Assuming 32 syndrome bits per possible combination as suggested by \cite{floresImplementationEvaluationSmart2023}, this leads to $\sim$88TB of storage space required for storing all syndromes without an index.}. This is mitigated by all approaches by assuming $N_{ID} = N_{ID}^{cell}$, which generally does not hold when \textit{pdcch-DMRS-scramblingID} is configured for the UE (see TS 38.211 \cite[Section 7.3.2.3]{ts38_211}). Secondly, error correction, which is required when the attacker does not have excellent signal quality, is only performed by Richards et al. \cite{richardsRNTIRecoveryOptimization2024} by leveraging hamming distances and longest common substrings across syndromes. Lastly, none of the approaches support common search space (CSS). We thus conclude that challenges 2 and 3 (unpredictability of channel and DCI descrambling) are not sufficiently solved.

Finding possible PDCCH configurations (challenge 1) is not investigated by these approaches. Mitigation of the approaches is possible by using either common search space (CSS) or configuring \textit{pdcch-DMRS-scramblingID} to be different than $N_{ID}^{cell}$.

\textbf{RACH side-channel} The second category of approaches \cite{wanNRScopePractical5G2024, luoSni5GectPracticalApproach, wangDownlinkControlInformation2026} intercept the initial RACH handshake between UE and gNB, obtaining the C-RNTI from message 4 (Contention Resolution) and, if provided, \textit{pdcch-DMRS-scramblingID} for USS as well as initial PDCCH configuration from RRC (RRCSetup). Subsequently, they attempt to decode candidates with the provided input parameters.

These approaches provide effective sniffing without requiring further brute force than the set of active C-RNTIs in the cell. However, they can only capture UEs that were detected during RACH and configurations indicated by SIB1 or disclosed during RACH before enabling encryption through AS security. Therefore, UEs already connected before activation of the sniffer and UEs where the handshake was not intercepted can not be captured. Furthermore, reconfiguring C-RNTIs and PDCCH configurations (via \textit{RrcReconfiguration}) after enabling AS security mitigates these approaches. This implies that all challenges towards performing passive sniffing are only solved given the previously mentioned restrictions.

\textbf{Divide and conquer} Ludant et al. \cite{ludant5GSniffingHarvesting2023} propose an approach which exploits identifier reuse in PDCCH DMRS, called \textit{5GSniffer}. The DMRS sequence used for channel estimation in PDCCH is initialized by only known factors (i.e. symbol index) and the scrambling factor $N_{ID}$ also used in scrambling of the DCI payload data. They thus first correlate all possible $2^{16}$ DMRS sequences at the given position (challenge 2) before brute-forcing DCI decoding with the remaining $2^{16}$ C-RNTI values, applying some optimizations such as attempting recently seen $N_{ID}$ and C-RNTI first. Configurations (challenge 1) are suggested to be obtained by the attacker beforehand, attaching to the target cell first and exporting relevant configuration parameters, requiring at least some active transmissions being sent by the attacker or following a RACH side-channel approach from above. Ludant et al. \cite{ludant5GSniffingHarvesting2023} propose to achieve configuration finding using brute-force over correlations on DMRS sequences. In case of highly interleaved CORESET configurations (e.g. 3 symbols in time and REG bundle size of 3), a brute-force approach can only rely on correlation of 6 DMRS bit per active group, not knowing the relations between active groups, while attempting $2^{16}$ possible scrambling sequences. Therefore for sufficiently interleaved CORESET configurations, their approach will result in a vast amount of false positives. Furthermore the grid (per carrier) can support up to 273 RBs in its maximum configuration \cite[Section 5.3.2]{ts38_104}, resulting in extremely large brute-force spaces being possible.

Importantly, cracking a DCI still requires a substantial amount of brute-force effort (challenge 3), increasing even further if this approach were to be extended towards automatically finding configurations (challenge 1) via constrained brute-force search. 3GPP could mitigate this approach by increasing the bit-length of $N_{ID}$ and C-RNTI or use a different factor for $N_{ID}$ in DMRS. Furthermore, gNBs could randomize PDCCH configurations per UE, even though this complicates scheduling.

We compared $\textit{5GSniffer}$ to our approach on their provided sample \textit{srsRAN 5GSniffer 1} and configuration, with RNTI range set to the full C-RNTI range, $N_{ID}$ restricted to $[1; 500]$ and all aggregation layers enabled. This sample with restricted settings represents the only sample and pseudo-realistic setting combination we could successfully process using \textit{5GSniffer} without it segfaulting, hinting at likely implementation issues. On this limited comparison, using our approach, multi-core processing time reduces $\approx$40$\times$ and total sample processing time reduces by $\approx$52\% from 7.79 seconds to 3.77 seconds on our reference machine$^{\ref{fn:reference_machine}}$. Furthermore, our approach found $\approx$17\% more valid DCI in the sample. This indicates some heuristic failures in \textit{5GSniffer}, which occur due to thresholds in the configuration provided by the authors not being properly fine-tuned to the sample. After fine-tuning thresholds manually, \textit{5GSniffer} also decodes all DCI on the given sample correctly, but processing times increase $\approx$7$\times$ to 49 seconds for the 5 seconds sample, loosing real-time capabilities. This corresponds to $\approx$300$\times$ of the multi-core processing time of our technique. The range for the scrambling factor $N_{ID}$ was only restricted for \textit{5GSniffer}. Similarly, only \textit{5GSniffer} was provided with the necessary configurations to parse the sample, both positioning \textit{5GSniffer} at a significant advantage. Apart from this, the experimental setup, including the resource limitations, were identical. This shows than on the singular sample that could be compared, \textit{5GSniffer} demonstrated significantly lower performance than our solution.

Therefore there currently exist no solutions addressing all three challenges simultaneously without significant restrictions, or requiring specification-level changes for mitigation. Especially, direct and efficient descrambling and decoding of binary DCI without using either brute-force or known parameters through side-channels has been a significant challenge \cite{luoSni5GectPracticalApproach}.

We do note however, that compared to related work, we do not extend further into post-processing of binary DCI contents, parsing the scheduled transmissions or subsequent attacks on privacy, which has been the case for some of the aforementioned related work. We provide an overview over related work and their capabilities in Table \ref{tab:capabilities}. Our work provides binary DCI contents, which can be post-processed to parse DCI and decode scheduled messages. We discussed these post-processing aspects but did not implement them. When message contents are known, UE-state such as the connection state can be reconstructed. Most subsequent attacks on privacy presented in related work then require the attacker to become active to, for example, send identifying traffic patterns. Examples included are flow linking across sessions, tracking of a user through a permanent identifier or engineered traffic patterns, and traffic classification of aspects such as user activity. Continuous live operation was demonstrated or claimed by all approaches. While some works were tested in real-world operational commercial networks, we did no such evaluation due to legal reasons.

\section{Conclusion}
In this paper we proposed a novel approach which facilitates obtaining the parameters and the binary contents of scrambled DCI candidates efficiently, including error correction, solving the challenge of DCI descrambling in $O(1)$ decoding attempts per candidate. Decoding DCI succeeds with below $1\%$ block error rate at low SNRs of $6.5$dB or higher. Furthermore, compared to previous comparable state of the art on a sample provided by them, our technique performed either $\approx$40$\times$ faster and obtained $\approx17\%$ more DCI when not fine-tuning thresholds for the previous state of the art, or 300$\times$ faster when fine-tuning thresholds to not miss DCI.
For the challenge of finding configurations, we extended the decoding approach with automatic finding of configurations using an incremental, constrained search, proving to be effective on samples captured from multiple base station implementations and configurations.
Furthermore, for the challenge of channel unpredictability in the absence of known reference symbols, we implemented active QPSK group detection and phase estimation, which consistently performs with $99\%$ accuracy above $10$dB SNR in our experiments. As a positive side-effect, this also reduces the search space size for configuration finding and binary DCI decoding. Lastly we implemented a full 5G synchronization and PDCCH descrambling pipeline utilizing our decoding approaches and evaluated it against various base station implementations and configurations with success. We make our open-source end-to-end binary DCI sniffer \textit{5GDescrambler} available to the public, alongside a small dataset of samples used in our evaluation.
 
Crucially, in order to mitigate our technique, the algebraic properties utilized need to be eliminated, requiring fundamental changes to the physical layer, such as replacing the scrambling function with a cryptographically sound non-linear function, thus breaking 3GPP specifications.
This shows that scheduling information in 5G is currently unprotected, which is concerning given that several studies have shown that leaking scheduling information on lower layers bears a significant privacy risk to users of a cell \cite{ludant5GSniffingHarvesting2023, wangWhatAppAppUsage, kohlsLostTrafficEncryption2019, baeWatchingWatchersPractical}. Due to the large deployment area of 5G, changes to lower-layer details, such as the scrambling function or RACH, are difficult. We would thus like to advocate for protection of PDCCH DCI, scheduling metadata, and other lower-layer channels and protocols for future generations of mobile networks, such as 6G.

As part of this work we disclosed our findings to GSMA and the vendors used in evaluation.
\subsection{Limitations and Future Work}
\label{sec:limitations}
We finally state some limitations of our approach and present some future work.

\textbf{RNTI LSB ambiguity in CSS} In CSS, when a candidate is rotated by 180 degrees (bit inversion), the LSB of RNTI flips. This can be mitigated by leveraging demodulation reference signals (DMRS) sent alongside PDCCH transmissions given the determined scrambling factor used in their construction (side-channel), estimation on subsequent PDSCH DMRS, or by using closely positioned CSI-RS. The closer the estimation is performed in time, the higher the expected rate of success. In case the last RNTI bit is required in traffic classification, performing one of these methods should be considered.

\textbf{Real-world measurement campaign} Due to legal reasons, we were not able to conduct a measurement campaign against local network operators. Our local authority, when asked for permission, stated that DCI scrambling can be seen as a lightweight concealment between operator and user, and as such may not be bypassed without this being considered eavesdropping on concealed communication. Since data collection would also affect users of the cell at that time, a one-sided contract with the operator was also not deemed sufficient. Therefore we evaluated on samples generated in MATLAB, captured from srsRAN and OpenAirInterface5G, as well as samples of commercial base stations provided by [state authority redacted]. Our approach targets the underlying specifications and performs well on the tested samples. Outside of this scope this remains unvalidated, since realistic scheduler load of real-world cells, including cells with larger bandwidths, vendor-specific behavior, mobility, handovers, untested cell configurations, or RF effects such as interference or impairments could affect results, including real-time capabilities.

\textbf{Pipeline parallelization} In our current pipeline implementation, candidate decoding waits for results from configuration finding in previous slots, leading to delays for candidate decoding when configuration finding stalls. Furthermore, when processing from SDR and configuration finding stalls for extended periods of time, queues of the pipeline may overflow, causing the pipeline to skip samples, missing the corresponding DCI. Currently this can be mitigated by disabling configuration finding entirely or after a set amount of frames. In future work we will parallelize the pipeline further, including scaling horizontally.

\textbf{Pipeline extension} \textit{5GDescrambler} currently implements a full 5G pipeline from cell synchronization to sniffing. For future work we want to extend the synchronization to support NSA synchronization, and extend the sniffer to interpret the received DCI bits in order to decode PUSCH/PDSCH, to determine the underlying encrypted traffic. Furthermore we want to utilize our sniffer to conduct traffic classification for identification of users and individual applications.

\begin{acks}  %
We would like to thank the Federal Office for Information Security (BSI), especially Team TEMIS and Mel Tellermann, for conducting experiments, sharing traces from commercial vendor setups for our study and artifact publication, and providing guidance during the disclosure process to affected parties.
This work has in part been funded by the Helmholtz Association through the KASTEL Security Research Labs (HGF Topic 46.23), by the German Research Foundation (DFG, Deutsche Forschungsgemeinschaft) as part of Germany’s Excellence Strategy – EXC 2050/1 – Project ID 390696704 – Cluster of Excellence \enquote{Centre for Tactile Internet with Human-in-the-Loop} (CeTI) of Technische Universität Dresden, and the Federal Ministry of Education and Research of Germany through project Open6GHub – Project ID 16KISK010.
This paper was checked for grammar and formulations using NotebookLM.
\end{acks}

\bibliographystyle{ACM-Reference-Format}
\balance
\bibliography{bibliography}

\appendix %

\section{Open Science} %

We provide a complete open-source implementation of the 5GDescrambler end-to-end binary DCI sniffer for 5G, written in Rust to ensure performance and reproducibility. The codebase implements cell synchronization, active group detection, our descrambling technique exploiting algebraic structure, constrained configuration inference, and optional visualization for validation.

All artifacts (source code, build instructions, dependencies, and configuration examples) are released publicly. The package further includes synthetic 5G RAN captures generated in controlled lab environments (OpenAirInterface5G, [vendor 1 redacted], [vendor 2 redacted]), and code reproducing the main evaluation results.

To address dual-use concerns, no captures from operational commercial networks are provided. The released artifacts nonetheless enable independent verification of all reported results, including decoding robustness and performance, configuration inference, active group detection, and feasibility analyses.

The artifact package follows the ACM reproducibility guidelines and includes documentation, test cases, and validation scripts to facilitate reuse, extension, and comparative evaluation by the community.\\

\noindent
All artifacts, including samples, are made available at:\\\url{\paperwebsite}

\section{Ethical Considerations} %

Our work demonstrates a novel technique enabling effective decoding of 5G PDCCH scheduling information, which carries significant potential for misuse including real-time user/device tracking, traffic analysis, and deanonymization.
We therefore treat this result as strictly security- and privacy-sensitive: operational details are limited to the minimum required to validate scientific claims, and artifact releases exclude any traces from operational networks or data exposing live users.

All experiments were conducted exclusively in controlled laboratory environments using open-source base stations (srsRAN, OpenAirInterface5G) and authorized test deployments ([vendor 1 redacted], [vendor 2 redacted]), with no collection of data from unsuspecting users. Reproducible samples use only synthetic or sanitized data to prevent direct misuse.

Publication serves a clear defensive purpose: to expose control channel design flaws, quantify their severity, and propose concrete mitigations.
Researchers replicating this work must observe all applicable interception, privacy, and radio regulations and restrict evaluations to isolated testbeds.

\textbf{Stakeholders and impacts} The primary stakeholders are mobile subscribers, private 5G users, network operators, equipment vendors, standards bodies, and the research community. If misused, our approach could provide enabling input for passive tracking, traffic analysis, and inference of sensitive user behavior, thereby harming user privacy and exposing operators and vendors to security and compliance risks.

\textbf{Research impact} This work presents a practically relevant technique exploiting algebraic structure to eavesdrop 5G scheduling information and shows that passive decoding can remain feasible even without prior side-channel leakage or active participation. The intended impact is defensive: to support risk assessment, guide mitigations, and inform protocol hardening and future standardization efforts.

\textbf{Mitigations} Because of the dual-use nature of the results, the presentation and release of artifacts should be restricted to what is necessary for scientific scrutiny. Experiments should be limited to controlled, authorized environments; released datasets should be synthetic or sanitized; and replication should be undertaken only in compliance with applicable radio, interception, and privacy laws. In parallel, responsible disclosure to affected vendors and relevant standardization bodies is necessary to reduce downstream harm.

\newpage
\nobalance
\section{Supplementary material}
\begin{minipagealg}
	\SetKwInOut{Input}{Input}
	\SetKwInOut{Output}{Output}
	
	\underline{function generateG} $(A,E,shared)$\;
	\Input{DCI payload size $A$, encoded DCI size $E$ and boolean value $shared$ indicating whether common search space is used}
	\Output{Generator matrix $G$}
	\tcp{K = |DCI|+|CRC| bits; N is polar kernel size}
	K = A + 24;
	N = getN(K, E)\;
	\tcp{Build and chain matrices for encoding steps}
	Mat<KxK> crc\_interleaver = permMat(getCRCInterleaverPattern(K))\;
	Mat<KxN> info\_bits = permMat(getInfoBitPattern(K, N, E))\;
	Mat<NxN> polar\_kernel = generatePolarMatrix(N)\;
	Mat<NxE> rate\_matching = permMat(getRateMatchingPattern(K, N, E))\;
	Mat<KxE> $P$ = crc\_interleaver * info\_bits * polar\_kernel * rate\_matching\;
	\tcp{CRC application to input vector $ones(24)||DCI$}
	Mat<KxK> crc = getCombinedCRCMatrix(A, K)\;
	\tcc{Input vector $u$ should contain K bits plus the negation bit, $RNTI$ and $N_{ID}$ bits (32).}
	U = K + 33\;
	\tcc{Create matrix masking last 16 CRC bits after DCI payload with $RNTI$ given in $u$ (offset 1 after K bits input).}
	Mat<33xK> rnti\_mask = moveMat(1, A + 8, 16)\;
	\tcp{XOR rnti to CRC by concatenating rows}
	Mat<UxK> apply\_crc = concatRows(crc, rnti\_mask)\;
	\tcp{Preserve negation bit, $RNTI$ and $N_{ID}$ for scrambling by appending them.}
	Mat<Ux33> append\_init = moveMat(K, 0, 33)\;
	Mat<UxU> dci\_init = concatColumns(apply\_crc, append\_init)\;
	\tcp{Use values copied by moveMat for scrambling}
	Mat<33xE> scramble = goldSequenceMatrix(E, shared)\;
	Mat<UxE> enc\_and\_scramble = concatRows($P$, scramble)\;
	\tcp{Chain initialization with remaining steps}
	Mat<UxE> $G$ = dci\_init * enc\_and\_scramble\;
	return $G$;
	
	\vspace{5px}
	function permMat $(R)$\;
	\For{$i = 0; i<|R|; i = i+1$}{
		$M[i, R[i]]$ = 1 \tcp*{Interleave bit index $i$ to $R[i]$}
	}
	return M\;
	
	\vspace{5px}
	function moveMat $(from, to, length)$\;
	\For{$i = 0; i<length; i = i+1$}{
		$M[from + i, to + i]$ = 1 \tcp*{Move range over}
	}
	return M\;
	
	\caption{Generation of $G$. 3GPP provided functions and concatenation functions are left out for brevity.}
	\label{alg:gen_matrix}
\end{minipagealg}

\begin{minipagealg}
	\SetKwInOut{Input}{Input}
	\SetKwInOut{Output}{Output}
	
	\underline{function correctErrorsAndDecode} $(C, K, E, shared, \hat{x})$\;
	\Input{The maximum amount of incorrect LLRs to consider $C$, DCI payload and CRC size $K$, encoded DCI size $E$, boolean value $shared$ indicating whether common search space is used, $\hat{x}$ the received vector as LLRs}
	\Output{Decoded vector $u$, or \textbf{null} if decoding failed}
	\tcp{Select likely error set from lowest $C$ LLRs}
	I = minIndexes($C$, $\hat{x}$)\;
	\tcp{Build a LES containing all known bits in $u$}
	known = [CRC,1,Padding]\;
	Mat<|known|xI> les = new Mat()\;
	Mat<|known|x1> solv = new Mat()\;
	\For{$j = 0; j<|known|; j=j+1$}{
		\For{$b = 0; b<|I|; b=b+1$}{
			\tcp{Copy contribution to u by invalid bit b}
			les[j, b] = $G^{-1}$[ind(j), ind(b)]
		}
		\tcp{Compute result of all non-invalid bits for equation}
		sum = 0\;
		\For{$b = 0; b<|\neg I|; b=b+1$}{
			\tcc{Bits set in $\hat{x}$ and the column in $G^{-1}$ contribute to output u in matrix multiplication}
			\If{$\hat{x}$(ind(b)) == 1 \textbf{and} $G^{-1}$[ind(j), ind(b)] == 1}
			{sum = sum $\oplus$ 1;}
		}
		\tcc{If expected value of known bit j is 1, we need to account for it}
		\If{knownValue(j) == 1}{sum = sum $\oplus$ 1;}
		solv[j, 0] = sum;
	}
	\tcp{Solve LES using gaussian elimination}
	gaussianElimination(les, solv)\;
	\tcc{Extract invalid bit bindings from LES, which is now in row-echelon form. In case multiple solutions exist, we extract one possible binding. If no solution exists we return 0.}
	$I_{corrected}$ = extractVariableResults(les, solv)\;
	\tcp{Correct vector and decode}
	e = $I \oplus I_{corrected}$\;
	x = $\hat{x} \oplus e$\;
	u = x * $G^{-1}$\;
	\tcp{Correct and check CRC, negation and padding bits}
	\textbf{if} ( $\neg$ correctAndMatchSyndrome(u) ) \{ return \textbf{null}; \}\\
	return u\;
	\caption{Error correction and decoding (simplified). \textit{ind} refers to the index of the respective bit in its origin vector $u$ or $\hat{x}$.}
	\label{alg:err_correction}
\end{minipagealg}

\begin{minipagealg}
	\SetKwInOut{Input}{Input}
	\SetKwInOut{Output}{Output}
	
	\underline{function determineActiveGroups} (frame, cutoff)\;
	\Input{The frame to operate on and a cutoff value determined by channel estimation}
	\Output{Groups considered active}
	active\_groups = []\;
	\For{symbol \textbf{in} frame}{
		\For{reg \textbf{in} symbol} {
			\tcp{Require at least 9 REs to be above cutoff}
			\If{$\neg$reg.count(|re| > cutoff) >= 9}{
				\tcp{Reduce groups by their std angle}
				angles = []\;
				\For{re \textbf{in} reg} {
					\tcc{Mirror RE to quadrant 1, spread out angle to full circle, recreate as complex number}
					angles.add(complexFromPolar(toQuadrant1(re).arg() * 4))\;
				}
				\If{std(angles) >= $\frac{\pi}{3}$} {
					continue\;
				}
				\tcp{Reduce groups by their std amplitude}
				factor = std(reg.norm()) / avg(reg.norm())\;
				\If{factor >= 0.3} {
					continue\;
				}
				\tcp{Else, consider group as active}
				active\_groups.add(reg)\;
			}
		}
	}
	\tcp{Rotate groups in a symbol so they match QPSK}
	\dots \\
	return active\_groups\;
	
	\caption{Active group detection}
	\label{alg:active_groups}
\end{minipagealg}

\begin{minipagealg}
	\SetKwInOut{Input}{Input}
	\SetKwInOut{Output}{Output}
	
	\underline{function processFrame} (active\_group\_frame, config)\;
	\Input{The groups considered active per frame, and a config}
	\Output{List of DCI in the frame and an extended config}
	dci = []\;
	\tcc{Attempt to decode all known candidates from config}
	\For{slot \textbf{in} active\_group\_frame}{
		\For{candidate \textbf{in} config}{
			\tcc{Attempt decoding if candidate is in active groups of frame}
			\If{slot.isActive(candidate)} {
				res = correctErrorsAndDecode(C, candidate.K, candidate.E, candidate.shared, slot.extract(candidate))\;
				\If{res != \textbf{null}} {
					\tcc{Add DCI to return set and exclude from further search}
					dci.add(res)\;
					slot.markInactive(candidate)\;
				}
			}
		}
	}
	\tcp{Attempt to find new configurations}
	\For{slot \textbf{in} active\_group\_frame}{
		\tcc{Attempt likely configurations according to our heuristic}
		\For{combination \textbf{in} slot.activeGroups().validCombinations()} {
			\tcc{Iterate all valid configurations of K, E and shared for the combination}
			\For{candidate \textbf{in} combination.validConfigurations()} {
				res = correctErrorsAndDecode(C, candidate.K, candidate.E, candidate.shared, slot.extract(candidate))\;
				\If{res != \textbf{null}} {
					\tcc{Add DCI to return set and exclude from further search}
					dci.add(res)\;
					slot.markInactive(candidate)\;
					\tcc{Add candidate to config, if not already present}
					\If{$\neg$ config.contains(candidate)} {
						config.add(candidate)\;
					}
				}
			}
		}
	}

	return dci, config\;
	
	\caption{Sniffing and configuration finding}
	\label{alg:sniffing_config}
\end{minipagealg}

\begin{table*}[t]
	\centering
	\begin{tabular}{|c|c|c|c|c|c|c|c|c|c|c|}
		\hline
		& $A$       & $K$       & $E$  & $N$ & $dim(G)$     & $rank(G)$ & $nullity(G)$ & $dim(G^{-1})$ & $rank(G^{-1})$ & $nullity(G^{-1})$ \\
		\hline
		\parbox[t]{2mm}{\multirow{7}{*}{\rotatebox[origin=c]{90}{UE-specific}}} & $[12; 44]$  & $[36; 68]$  & 108  & 128 & (57 + A)x108 & 56 + A    & 1            & 108x108       & 108            & 0                 \\
		& $[45; 52]$  & $[69; 76]$  & 108  & 128 & (57 + A)x108 & 55 + A    & 2            & 108x108       & 108            & 0                 \\
		& $[53; 83]$  & $[77; 107]$ & 108  & 128 & (57 + A)x108 & 108       & A - 51       & 108x108       & 108            & 0                 \\
		& $[12; 140]$ & $[36; 164]$ & 216  & 256 & (57 + A)x216 & 56 + A    & 1            & 216x216       & 216            & 0                 \\
		& $[12; 140]$ & $[36; 164]$ & 432  & 512 & (57 + A)x432 & 56 + A    & 1            & 432x432       & 432            & 0                 \\
		& $[12; 140]$ & $[36; 164]$ & 864  & 512 & (57 + A)x864 & 56 + A    & 1            & 864x864       & 512            & 352               \\
		& $[12; 140]$ & $[36; 68]$  & 1728 & 512 & (57 + A)x1728 & 56 + A    & 1            & 1728x1728     & 512            & 1216              \\
		\hline
		\parbox[t]{2mm}{\multirow{7}{*}{\rotatebox[origin=c]{90}{Shared}}} & $[12; 66]$  & $[36; 90]$  & 108  & 128 & (57 + A)x108 & 41 + A    & 16           & 108x108       & 108            & 0                 \\
		& 67        & 91        & 108  & 128 & (57 + A)x108 & 107       & 17           & 108x108       & 108            & 0                 \\
		& $[68; 83]$  & $[92; 107]$ & 108  & 128 & (57 + A)x108 & 108       & A - 51       & 108x108       & 108            & 0                 \\
		& $[12; 140]$ & $[36; 164]$ & 216  & 256 & (57 + A)x216 & 41 + A    & 16           & 216x216       & 216            & 0                 \\
		& $[12; 140]$ & $[36; 164]$ & 432  & 512 & (57 + A)x432 & 41 + A    & 16           & 432x432       & 432            & 0                 \\
		& $[12; 140]$ & $[36; 164]$ & 864  & 512 & (57 + A)x864 & 41 + A    & 16           & 864x864       & 512            & 352               \\
		& $[12; 140]$ & $[36; 164]$ & 1728 & 512 & (57 + A)x1728 & 41 + A    & 16           & 1728x1728     & 512            & 1216              \\
		
		\hline
	\end{tabular}
	\caption{Effect of chosen $A$/$K$ and $E$ on the properties of the resulting matrices $G$ and $G^{-1}$.}
	\label{tab:mat_rank}
\end{table*}

\end{document}